\documentclass[%
 preprint, 
 amsmath,amssymb,
 aps, physrev,
showkeys,
floatfix,
tightenlines,
]{revtex4-2}

\usepackage{graphicx}%
\usepackage{dcolumn}%
\usepackage{bm}%

\usepackage{booktabs}
\usepackage{url}
\usepackage{xcolor}
\usepackage{tikz}

\usepackage[hyphens]{xurl}
\usepackage{mathtools}
\usepackage{siunitx}
\usepackage[noabbrev]{cleveref}
\usepackage[vlined,algo2e]{algorithm2e}
\usepackage{listings}       %
\usepackage{caption}
\usepackage{subcaption}

\pgfkeys{/pgf/number format/.cd,sci,std={-1:1},sci generic={mantissa sep={\times},exponent={10^{#1}}}}

\def\num#1{\pgfkeys{/pgf/number format/.cd,std, precision=4,zerofill}\pgfmathprintnumber{#1}}

\def\rnum#1{\pgfkeys{/pgf/number format/.cd,std, precision=2, zerofill}\pgfmathprintnumber{#1}}

\usepackage{acro}
\DeclareAcronym{4DMRI}{
  short=4D PC-MRI,
  long=4D phase-contrast MRI,
}
\DeclareAcronym{3DMRI}{
  short=3D bSSFP MRI,
  long=3D balanced steady-state free precession (3D bSSFP) ,
}
\DeclareAcronym{CG1}{
  short=CG1,
  long=continuous piecewise linear,
}
\DeclareAcronym{DG0}{
  short=DG0,
  long=discontinous piecewise constant,
}
\DeclareAcronym{IP}{
  short=IP,
  long=interior penalty,
}

\newcommand{\vin}{\mathbf{v}_{\text{in}}}
\newcommand{\bndry}[1]{\Gamma_\text{#1}}

\begin{document}

\title{\textbf{%
\emph{In vivo} evidence of blood flow slippage: failure of the no-slip boundary condition assumption
}}%

\author{Alena~Jarolímová}
\author{Jaroslav~Hron}%
 \email{Contact author  (mathematical issues): \href{mailto:jaroslav.hron@matfyz.cuni.cz}{jaroslav.hron@matfyz.cuni.cz}}
 \author{Karel~Tůma}
\affiliation{Faculty of Mathematics and Physics, Charles University, Prague, Czech Republic}

\author{Radomír~Chabiniok}
\affiliation{Division of Pediatric Cardiology, Department of Pediatrics, UT Southwestern Medical Center, Dallas, TX, United States}

 \author{Josef~Málek}
\affiliation{Faculty of Mathematics and Physics, Charles University, Prague, Czech Republic}

\author{Keshava~Rajagopal}
 \email{Contact author (medical issues): \href{mailto:keshava.rajagopal@jefferson.edu}{keshava.rajagopal@jefferson.edu}}
\affiliation{%
Division of Cardiac Surgery, Department of Surgery, Sidney Kimmel Medical College of Thomas Jefferson University, Philadelphia, PA, United States}%

\collaboration{
In memory of our mentor and teacher Kumbakonam Ramamani~Rajagopal \\
\textborn~November 24, 1950 -- \textdied~March 20, 2025
}\noaffiliation

\begin{abstract}
The assumption that blood adheres to vessel walls, the ``no-slip'' boundary condition, is an essential premise of cardiovascular fluid dynamics. Yet, whether it holds true \emph{in vivo} has not been established. Using 4D flow magnetic resonance imaging of the human thoracic aorta and modeling blood as a Navier--Stokes fluid, we quantify the velocity of blood at the wall. We find tangential wall velocities of about 30--80\% of the mean luminal velocity, providing clear evidence of blood slippage. To our knowledge, this is the first demonstration that the no-slip condition does not apply to blood flow \emph{in vivo}. This finding challenges a fundamental assumption in cardiovascular modeling and directly affects key blood flow characteristics such as pressure drop, vorticity, wall shear stress, and energy dissipation, all of which play important roles across a wide range of cardiovascular conditions.
\end{abstract}


\maketitle

\section{Introduction}
In simulations and other theoretical studies of blood flow, it is commonly assumed that blood adheres to the vessel walls, described by a well-known `no-slip' boundary condition \cite{DARRIGOL2002,Nichols2022,Raissi2020,Stokes1845}. However, there are several reasons that suggest that assumption of the no-slip condition may be unfounded \cite{Nolte2019,nubar,Raissi2020,Yilmaz:2008uj}. These include: (i) variable aortic surface and wall properties, (ii) the complex composition of blood as a heterogeneous mixture, and (iii) the fact that the total dissipation can vary with the slip parameter in a non-monotonic way \cite{Chabiniok2022}. Numerous studies have focused on the occurrence of slip in specific situations (see, e.g., \cite{Craig2001,Neto2005,Vinogradova2006} for experimental demonstrations of boundary slip under controlled shear and surface-chemistry conditions, the reviews \cite{Denn2004,Hervet2003} for broader discussions spanning simple and complex fluids and the role of surface roughness and wettability, and the recent methodological works \cite{MR1,MR2} for systematic frameworks to determine whether and to what extent a fluid slips adjacent to a solid surface).

This study provides evidence arguing against the validity of the no-slip condition. Specifically, we use magnetic resonance imaging (MRI) data to demonstrate that blood slips against the arterial wall \emph{in vivo}. The presence of wall slip significantly changes flow dynamics, as reflected in key hemodynamic parameters \cite{Chabiniok2022,Fara2024}. These include the pressure gradient, fluid vorticity, and wall shear stress. Since typical MRI velocity data do not possess high resolution, we aim to improve boundary velocity information by applying data assimilation to the flow model, leveraging all available information from the entire volume of interest.

\section{Methods}
We used seven different sets of MRI data of healthy volunteers covering the descending thoracic aorta. The datasets were acquired previously at King’s College London \cite{Marlevi2021}. Each dataset includes both the vascular geometry (3D balanced steady-state free precession) and the velocity field (4D phase-contrast MRI with temporal resolution of $\approx 35$ ms, acquired with 8-fold acceleration and reconstructed using the k-t PCA technique \cite{kt-PCA2009} combined with a sparsifying transform \cite{kt-PCA2012}). Our aim is to identify the slip parameter at the boundary (allowing no-slip as a possibility) that best reproduces the measured velocity fields. We assume the following boundary conditions on the wall:
\begin{align}
    \mathbf{v}\cdot\mathbf{n} = 0, \qquad
    \kappa\,\mathbf{v}_{\mathbf{\tau}} = -(\mathbb{T}\mathbf{n})_\mathbf{\tau}, \label{p:kappa}
\end{align}
where the first equation represents the impermeability condition, while the second one relates the tangential velocity $\mathbf{v}_{\tau} := \mathbf{v} - (\mathbf{v}\cdot\mathbf{n})\mathbf{n}$ to the wall shear stress $(\mathbb{T}\mathbf{n}){\boldsymbol{\tau}}$, that is, the tangential component of the normal traction $\mathbb{T}\mathbf{n}$. Here, the symbol $\mathbb{T}$ denotes the Cauchy stress tensor and $\mathbf{n}$ is the unit outward normal to the wall.
The extent of slippage is controlled by the non-negative scalar parameter $\kappa$. The case $\kappa = 0$ corresponds to perfect slip, while $\kappa>0$ characterizes Navier's slip;
the no-slip condition is obtained by permitting $\kappa$ tend to infinity.

\medskip

\subsection{Governing equations}  

We consider unsteady blood flow in a three-dimensional domain $\Omega$ representing a part of the descending thoracic aorta. Based upon \cite{fung1997biomechanics, ku1997blood, formaggia2009cardiovascular, berger2000stenotic, Nichols2022}, it is reasonable to model blood as a Navier–Stokes fluid, particularly in the context of aortic hemodynamics. Moreover, in large arteries and under normal flow regimes---such as those examined in baseline studies aimed at assessing blood slippage along the endothelium---blood flow is generally modeled as being non-turbulent, see also \cite{Scarselli2023} for further arguments supporting this in the context of pulsatile flows.

Hence, the governing system of partial differential equations is that of an incompressible Newtonian fluid, described by the incompressible Navier–Stokes equations:
\begin{align}\label{eq:NS}
        &&\mathrm{div}\,\mathbf{v} &= 0,\quad &&&&\text{in } (0,T)\times \Omega, \\ 
        &&\rho\, \partial_t\mathbf{v} + \rho(\nabla\mathbf{v})\mathbf{v} &= \mathrm{div}\,\mathbb{T}, \quad &&&&\text{in } (0,T)\times \Omega, \\
        &&\mathbb{T} &= -p\,\mathbb{I} + 2\mu\,\mathbb{D}(\mathbf{v}), \quad  &&&&\text{in } (0,T)\times \Omega.
\end{align}
Here, $T > 0$ is the terminal time of the period of our interest, $\mathbf{v}$ is the fluid velocity, $p$ the pressure, $\mathbb{T}$ the Cauchy stress tensor, $\mu$ the dynamic viscosity
and $\mathbb{D}(\mathbf{v})=\frac12(\nabla\mathbf{v}+(\nabla\mathbf{v})^{\rm T})$ denotes the symmetric part of the velocity gradient. Finally, $\rho$ stands for a positive constant density.

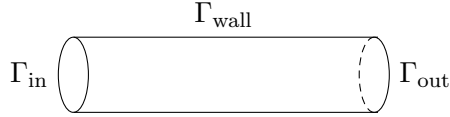
\begin{figure}
    \centering
\begin{tikzpicture}[scale=0.5]
\begin{scope}
  \coordinate (A) at (0mm,10mm);
  \coordinate (B) at (80mm,10mm);
  \coordinate (C) at (80mm,-10mm);
  \coordinate (D) at (0mm,-10mm);
  \draw [] (A) -- node[midway, above] {$\bndry{wall}$} (B) -- ++(1mm,0mm);
  \draw [] (D) -- (C) -- ++(1mm,0mm);
  \draw (B) arc [x radius = 4mm, y radius = 10mm, start angle=90, end angle=-90] node[midway, right] {$\bndry{out}$} (C);
  \draw (D) arc [x radius =-4mm, y radius = 10mm, start angle=-90, end angle=90] node[midway, left] {$\bndry{in}$} (A);
  \draw (A) arc [x radius =4mm, y radius = 10mm, start angle=90, end angle=-90] (D);
\draw [densely dashed] (C) arc [x radius = -4mm, y radius = 10mm, start angle=-90, end angle=90] (B);
\end{scope}
\end{tikzpicture}
    \caption{A scheme of a simple vessel with marked boundary types.}
    \label{fig:geometry}
\end{figure}

The boundary $\partial\Omega$ is divided into the inlet $\bndry{in}$, outlet $\bndry{out}$, and vessel wall $\bndry{wall}$,
each associated with distinct boundary conditions in the model, see Figure \ref{fig:geometry}. While the inlet and outlet are artificial boundaries introduced to define the domain of interest, $\bndry{wall}$ represents the interface between the flow domain and the vessel wall. 

On $\bndry{wall}$, we impose an impermeability condition (i.e., no flow in the normal direction) while allowing the fluid to slip along the wall in the tangential directions. The extent of this slip ranges from perfect slip (a frictionless wall), through Navier's slip (where the tangential velocity is proportional to the projection of the traction onto the tangent plane), to no-slip (where the fluid adheres completely to the boundary). All these scenarios are captured by the slip parameter $\kappa$, which may vary in space and represents the degree of friction: $\kappa = 0$ corresponds to a frictionless surface (full slip), while $\kappa = +\infty$ corresponds to the no-slip. At the inlet, inflow velocity $\vin$ needs to be prescribed. At the outlet, provided it is chosen appropriately, we assume that the flow is purely normal to the cross-section, and we impose a ``directional do-nothing'' condition to stabilize a possible backflow.
Boundary conditions are imposed as
\begin{align}
&&\mathbf{v}&=\mathbf{v}_{\rm in},
    &&&&\text{on } (0,T)\times \bndry{in}, \\
    (\mathbb{T}\mathbf{n})\cdot\mathbf{n} &= \tfrac{1}{2}\rho(\mathbf{v}\cdot\mathbf{n})_-^2, 
    \quad &\mathbf{v}_\tau &= 0,  &&&&\text{on } (0,T)\times \bndry{out}, \\
    \mathbf{v}\cdot\mathbf{n} &= 0, \qquad &
    \kappa \mathbf{v}_\tau &= -(\mathbb{T}\mathbf{n})_\tau , 
     &&&&\text{on } (0,T)\times \bndry{wall}.\label{Eq:bcs}
\end{align}
Importantly, the slip parameter $\kappa$ must be non-negative due to the physical requirement that the total dissipation resulting from fluid–wall interaction must be non-negative, in accordance with the Second Law of Thermodynamics. However, the constraint $\kappa > 0$ introduces certain limitations—such as the inapplicability of some numerical methods and computational tools. To address this, we introduce an auxiliary slip parameter $K$, which is allowed to take any real value in the range $(-\infty, +\infty)$, and relate it to $\kappa$ via the transformation:
\begin{equation}
    \label{Eq:2}    K = \ln\kappa \qquad \iff \qquad \kappa = \exp K.
\end{equation}

In practice, the inlet velocity $\vin$ and the wall slip parameter $\kappa$ cannot be measured directly and  cannot be then considered as known (given) quantities. Therefore, we treat them as unknowns, and determine their values based on the available 4D MRI data through an assimilation process, as developed in \cite{steady} and briefly described in the next subsection (some additional information is provided in the Supplementary Material). Our approach allows us to treat $\mathbf{v}_{\rm in}$ and $K$ as spatially varying.

\medskip

\subsection{Data assimilation and optimization.} 
The measured data are first processed by spatial interpolation to obtain their representation $\mathbf{d}_{\text{MRI}}$ on the computational mesh (see end of Section~1 in Supplementary Material for details). 
To measure the difference between the given velocity data $\mathbf{d}_{\text{MRI}}$ and the velocity field $\mathbf{v}=\mathbf{v}(\vin, K)$ obtained by the solution of \eqref{eq:NS}--\eqref{Eq:bcs} and viewed as the function of $\vin$ and $K$ that we wish to determine, we introduce the functional $\mathcal{J}$ as the $L^2$ norm of their difference, i.e., 
\begin{equation}\label{Eq:3}
    \begin{split}
    \mathcal{J}(\mathbf{v}(\vin,K)) = \frac{1}{2}\|\mathcal{M}(\mathbf{v}(\vin, K))-\mathbf{d}_{\text{MRI}}\|^2_{L^2(0, T, L^2(\Omega))} := \int_0^T \int_{\Omega} |\mathcal{M}(\mathbf{v})-\mathbf{d}_{\text{MRI}}|^2 \, \textrm{d}x\textrm{d}t,
    \end{split}
\end{equation}
where $\mathcal{M}$ is a measurement operator that maps the velocity field $\mathbf{v}$ to the same temporal 
resolution as $\mathbf{d}_{\text{MRI}}$, for details see Section~2 in Supplementary Material.  
The objective is to find $(\vin, K)$ that minimizes this error functional~$\mathcal{J}$. For more details on data assimilation see \cite{steady,chapelle2013fundamental,moireau2013sequential,marsden2014optimization,arostica2025parameter,manganotti2021coupling}.

To ensure that the optimization problem is well-posed, $\mathcal{J}$ is augmented with regularization terms $\mathcal{R}_1(\vin)$ and $\mathcal{R}_2(K)$, which are designed to enforce the smoothness of the estimated boundary parameters $\vin$ and $K$. Specifically, we include regularization terms that penalize excessive spatial and temporal variations in the inlet velocity profile $\vin$, as well as spatial irregularities in the auxiliary slip parameter $K$. %
These regularization terms play a crucial role in stabilizing the optimization problem by preventing overfitting to noisy data, and promoting physically realistic reconstructions of the boundary conditions.

With these regularization terms, the reduced error functional takes the form: %
\begin{align}\label{Eq:4}
    \begin{split}
        \mathcal{J}_R(\vin, K) :&= \mathcal{J}(\mathbf{v}(\vin,K)) + \mathcal{R}_1 (\vin) + \mathcal{R}_2 (K) \\ &=\frac{1}{2TL^3}\|\mathcal{M}(\mathbf{v}(\vin, K))-\mathbf{d}_{\text{MRI}}\|^2_{L^2(0, T, L^2(\Omega))} \\
        &\quad +\frac{\alpha}{2}\left(\frac{1}{T L^2}||\vin||^2_{L^2(0, T, L^2(\bndry{in}))} + \frac{1}{T}||\nabla\vin||^2_{L^2(0, T, L^2(\bndry{in}))} + \frac{T}{ L^2}||\partial_t{\mathbf{v}}_{\text{in}}||^2_{L^2(0, T, L^2(\bndry{in}))}\right)\\
        &\quad +\frac{\gamma}{2}||\nabla K||^2_{L^2(\bndry{wall})},
    \end{split}
\end{align}
where $T$ and $L$ are the terminal time and the characteristic length, ensuring the terms have the same units, and $\alpha$, $\gamma$ are regularization parameters, the values of which are specified in Table~S1 of the Supplementary Material.
We conclude this section by summarizing the objective of our study:
\begin{equation}
  \textrm{ To find } (\vin, K) \textrm{ that minimizes } \mathcal{J}_R \textrm{ whereas } \mathbf{v}(\vin, K) \textrm{ solves \eqref{eq:NS}--\eqref{Eq:bcs}.}\label{Eq:DA_method}
\end{equation}
This optimization method provides the desired value of $\kappa$  related to $K$ through \eqref{Eq:2}, denoted by $\kappa_{\rm opt}$, which is primarily treated as constant, although it can also be allowed to vary spatially. The method further determines the inlet velocity $\vin$, which is spatially varying.
To ascertain whether blood slips at the boundary, we also compute the average tangential velocity at the wall, and compare it to the average velocity within the entire domain. 

\section{MRI data and their segmentation}
The slip parameter, which is the result of our fitting process, is highly sensitive to the domain used for data fitting. If the segmented domain is smaller than the actual lumen, velocities farther from the boundary, closer to the center, are erroneously treated as boundary velocities, leading to an underestimated slip parameter $\kappa$. Conversely, if the domain is larger than the real lumen, velocities near zero (outside the lumen) are incorporated as boundary velocities, resulting in an overestimated slip parameter $\kappa$. Therefore, accurate segmentation of the lumen is crucial for reliable analysis.

\begin{figure}[!ht]
    \centering
\includegraphics[width=0.24\linewidth]{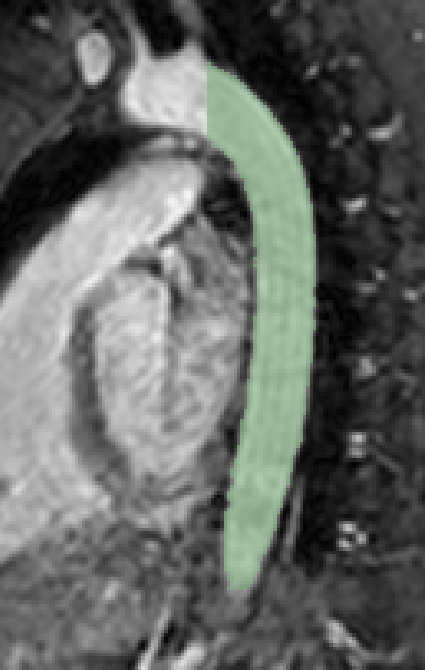}
    \quad
\includegraphics[width=0.24\linewidth]{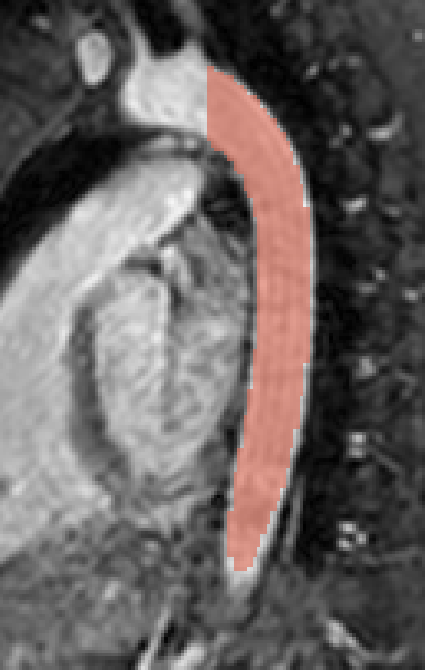}
    \quad
\includegraphics[width=0.24\linewidth]{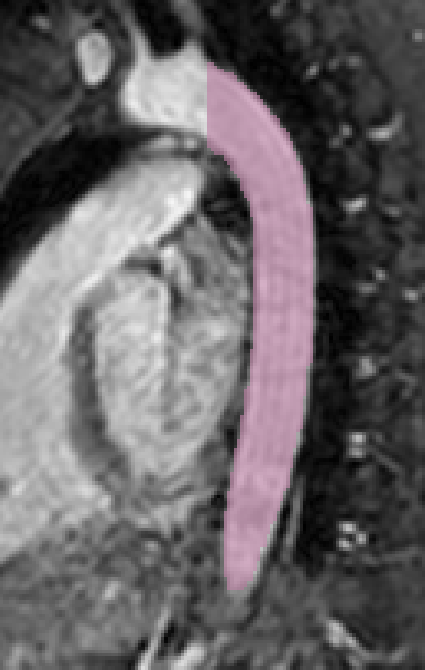}
\\[10pt]
\includegraphics[width=0.24\linewidth,trim={30mm 0 30mm 0},clip]{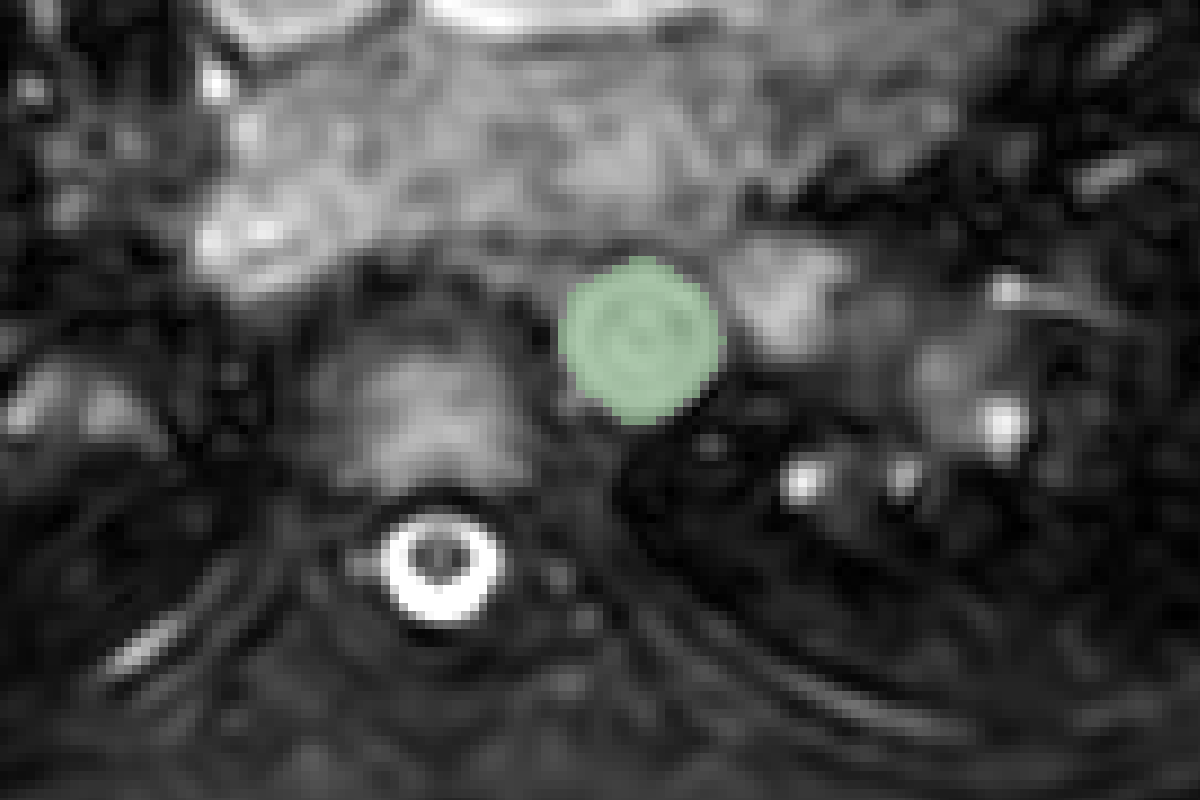}
    \quad
\includegraphics[width=0.24\linewidth,trim={30mm 0 30mm 0},clip]{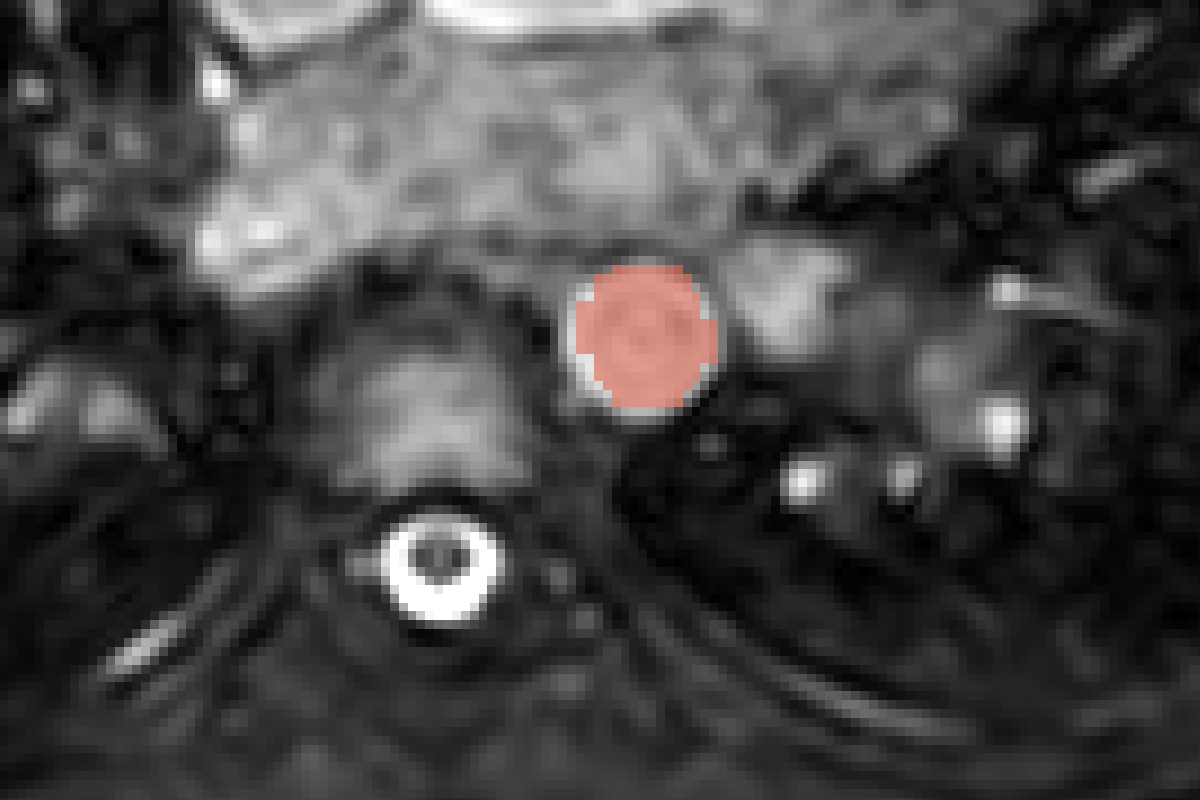}
    \quad
\includegraphics[width=0.24\linewidth,trim={30mm 0 30mm 0},clip]{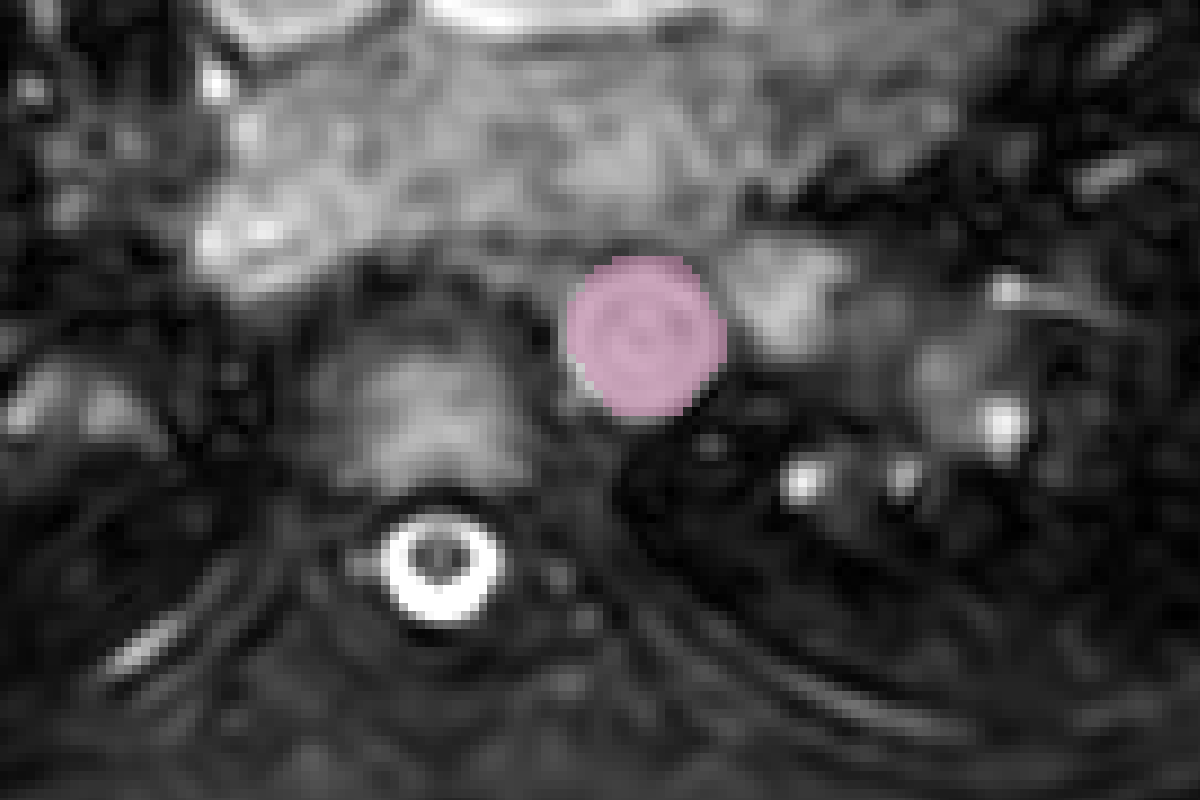}

    \caption{Comparison between the manual segmentation (left), automatic segmentation using Total Segmentator (middle) and automatic segmentation with an additional layer of pixels (right) on a cross-section in the middle of the geometry.}
    \label{fig: segmentation comparison}
\end{figure}

We segment the lumen using two methods: (i) manual segmentation of MRI data; (ii) automatic segmentation using artificial intelligence as implemented in TotalSegmentator \cite{Wasserthal2023}, which is trained on 1,204 CT and 298 MRI scans annotated by multiple researchers. The key advantage of the automatic approach is its consistency and reproducibility. Unlike manual segmentation, it eliminates operator bias, ensuring objective results. Notably, the domain obtained through automatic segmentation is slightly smaller than the manually segmented one (which results in a smaller slip parameter $\kappa$ for automatic segmentation). To address this, we also performed fitting with the domain generated by TotalSegmentator, and expanded it by one additional layer of pixels (1 px = 1.05 mm in our MRI data). Moreover, by adding an extra layer, the fitted slip parameter increases $\kappa$ – bringing the whole setting closer to the no-slip condition; if, even in this situation of increased $\kappa$ (relatively), significant slippage is observed, such a result underscores the validity of these findings. \Cref{fig: segmentation comparison} compares the manual and automatic segmentation methods.

\clearpage
\section{Results}
The results of our assimilation method presented above and applied to 4D~flow MRI datasets of the human thoracic aorta demonstrate that the classical no-slip boundary condition fails to accurately fit the measured velocity fields, revealing clear evidence of wall slip \emph{in vivo}. We then examine the physiological implications of this finding, showing that slip markedly reduces wall shear stress, an important example of a  key hemodynamic quantity.

\subsection{\emph{In vivo} failure of the no-slip condition}
\begin{figure}[!ht]
    \centering
    \includegraphics[width=0.3\linewidth]{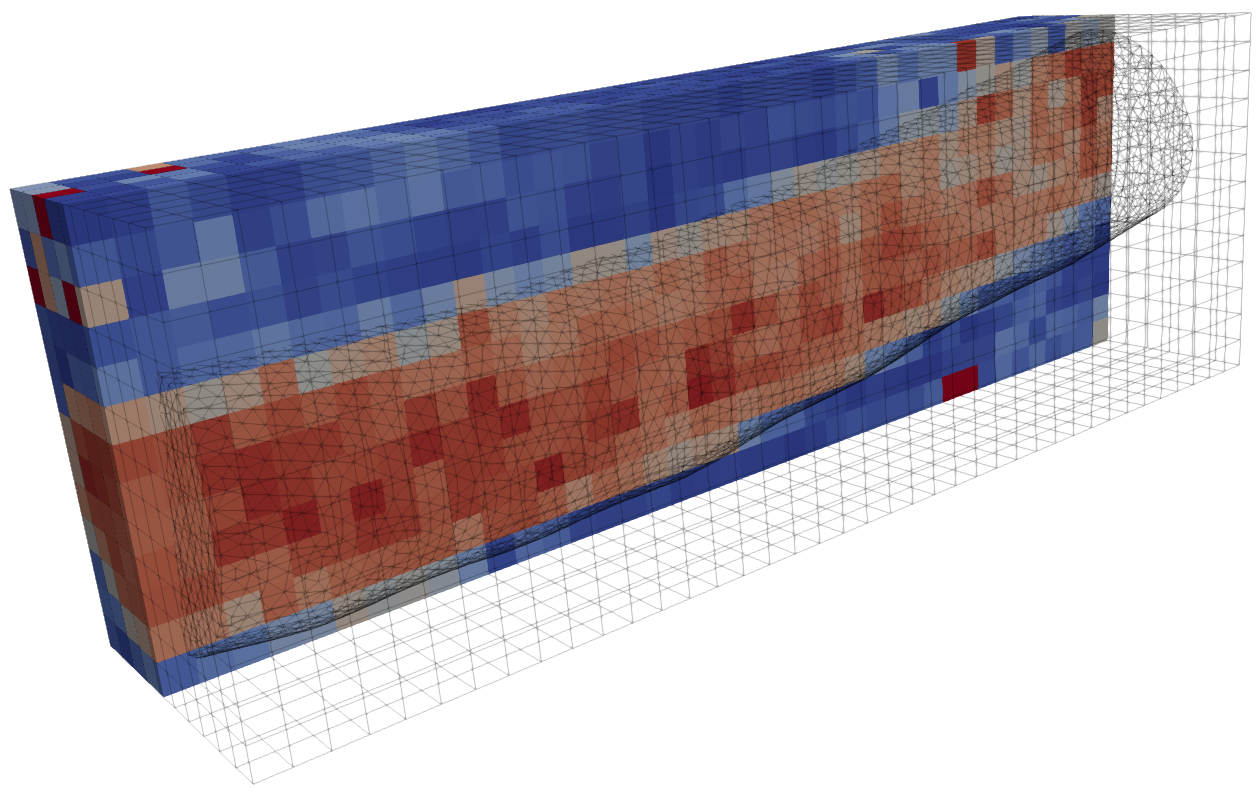} \hfill
    \includegraphics[width=0.3\linewidth]{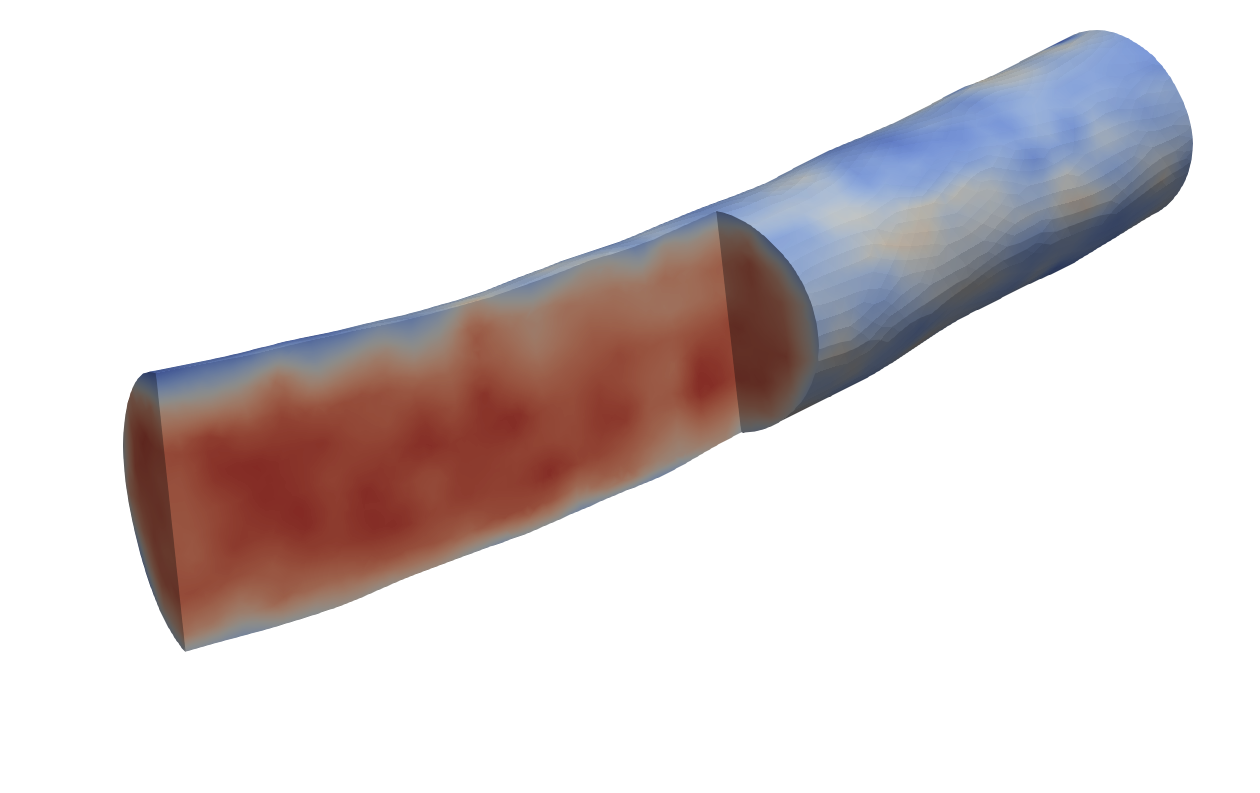} \hfill
    \includegraphics[width=0.3\linewidth]{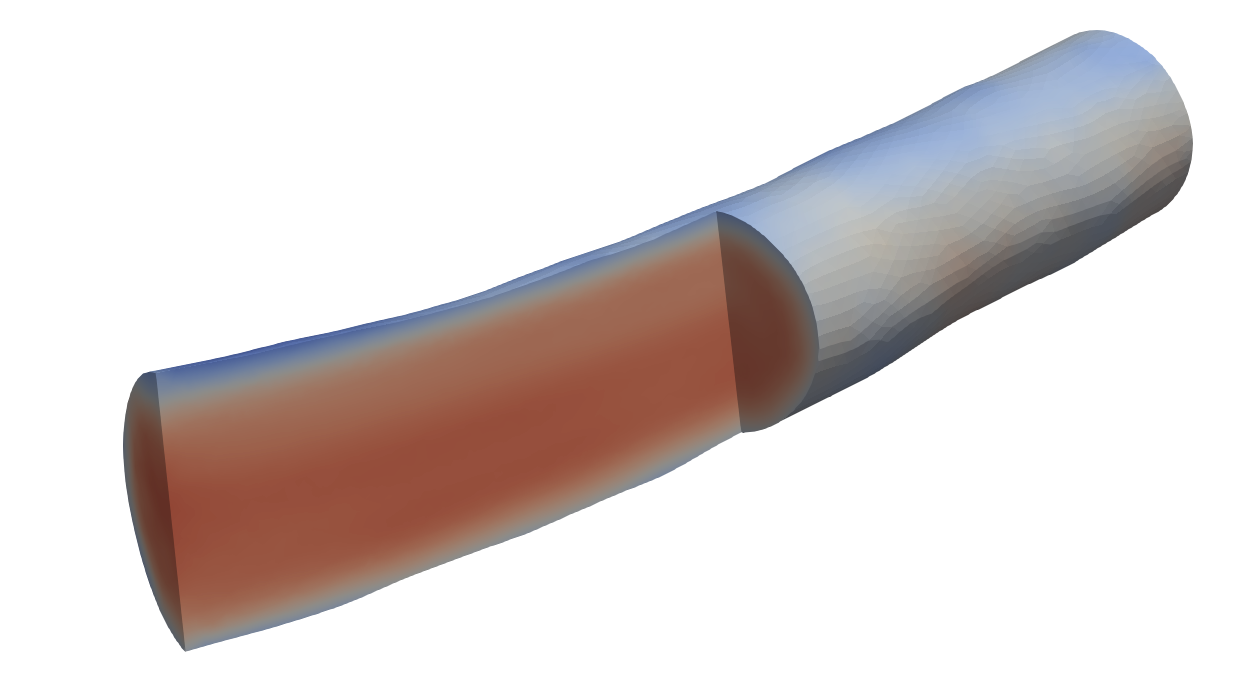}
    \caption{Comparison of the velocity fields for Dataset 1: The velocity field obtained from MRI data (left); the velocity field obtained from MRI data interpolated into a computational mesh, here based on manual segmentation method (middle); the velocity field obtained from numerical simulation with the optimal slip parameter $\kappa_{\rm opt} = \qty{3.89}{Pa.s/m}$ (right).}
    \label{fig: velocity comparison}
\end{figure}

A comparison between the velocity field obtained from MRI data and that from the numerical simulation is shown in \Cref{fig: velocity comparison}.

\begin{figure*}[!ht]
    \centering
    \footnotesize
\begin{tabular}{|c|c|c|}
\hline
Manual segmentation & Automatic segmentation & Automatic seg. + 1 voxel \\
\hline
\includegraphics[width=0.27\textwidth]{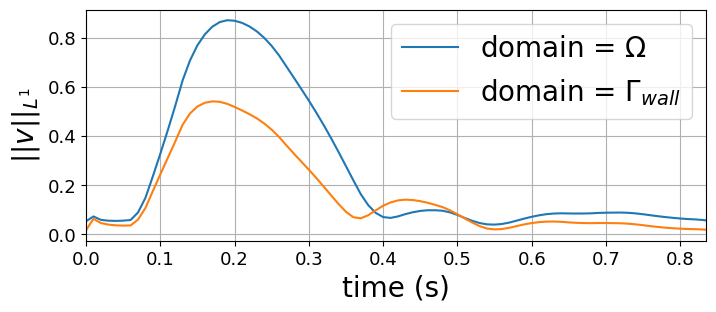} & \includegraphics[width=0.27\textwidth]{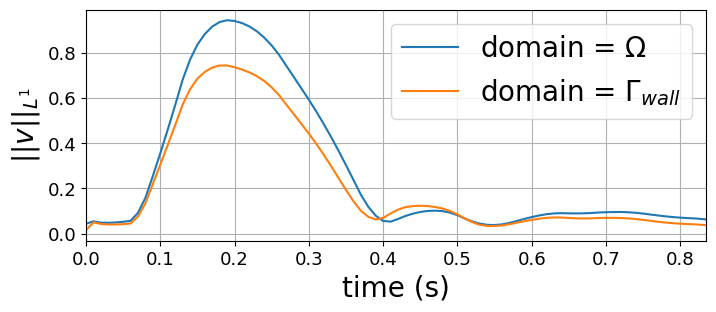} & \includegraphics[width=0.27\textwidth]{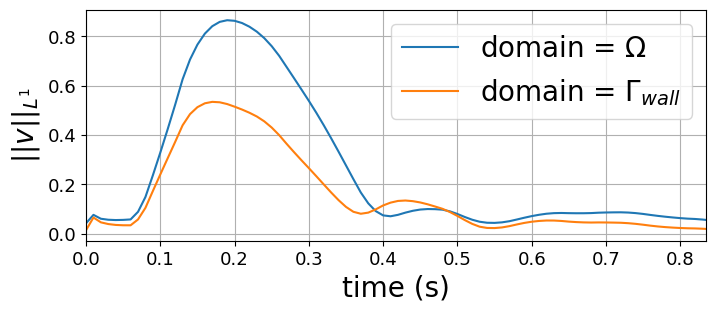} \\
$\rnum{61.910655046300725}\%$ & $\rnum{79.6258732112369}\%$ & $\rnum{61.92194314094098}\%$ \\
\hline
\includegraphics[width=0.3\textwidth]{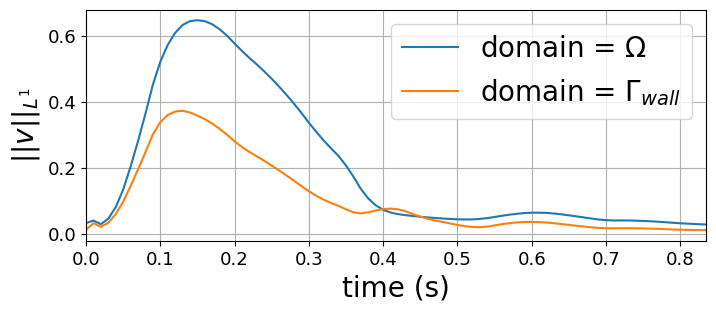} & \includegraphics[width=0.3\textwidth]{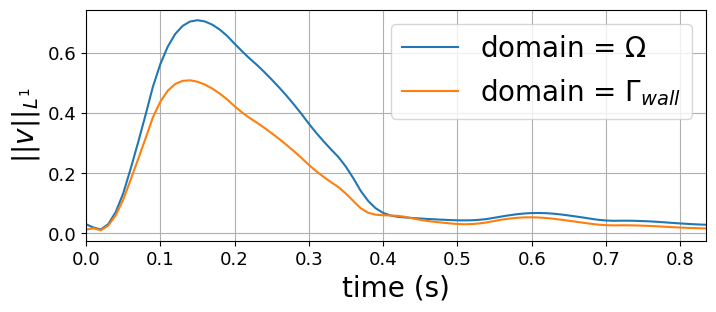} & \includegraphics[width=0.3\textwidth]{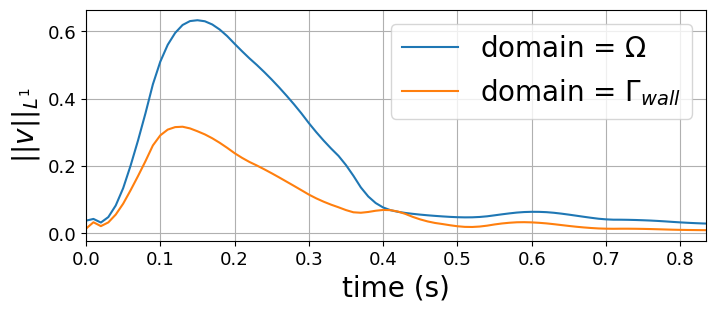} \\
$\rnum{53.126645197260395}\%$ & $\rnum{69.99737834150119}\%$ & $\rnum{46.54648036399473}\%$ \\
\hline
\includegraphics[width=0.3\textwidth]{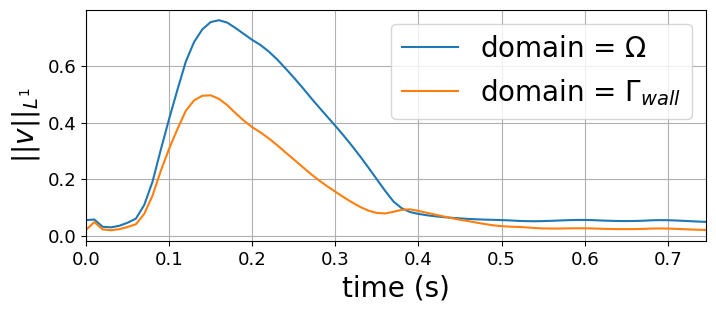} & \includegraphics[width=0.3\textwidth]{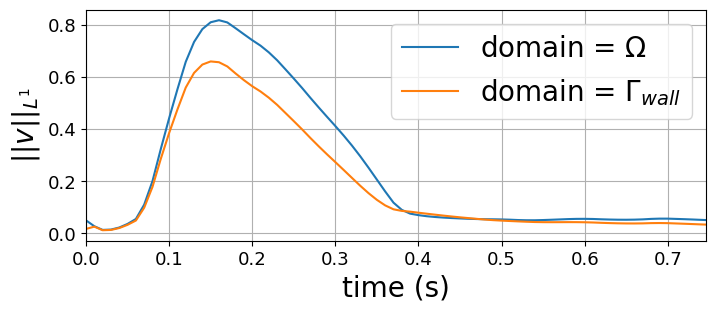} & \includegraphics[width=0.3\textwidth]{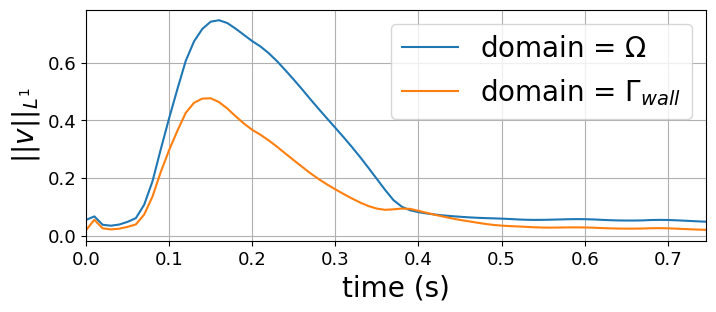}  \\ 
$\rnum{57.79473477818741}\%$ & $\rnum{77.68608611747857}\%$ & $\rnum{57.72266478420703}\%$ \\
\hline
\includegraphics[width=0.3\textwidth]{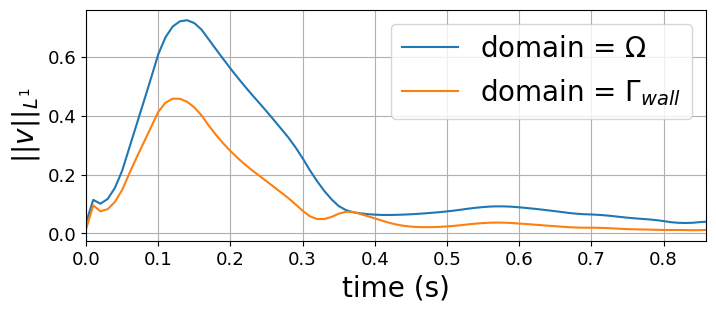} & \includegraphics[width=0.3\textwidth]{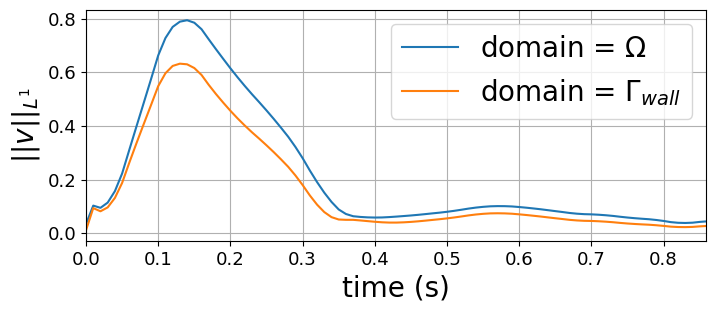} & \includegraphics[width=0.3\textwidth]{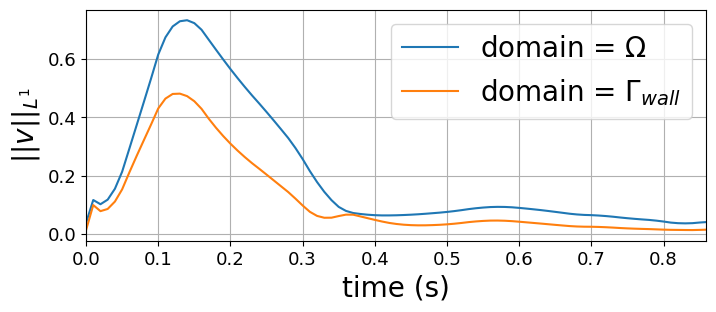} \\
$\rnum{52.45465401026156}\%$ & $\rnum{74.61633317016299}\%$ & $\rnum{56.954845675921604}\%$ \\
\hline
\includegraphics[width=0.3\textwidth]{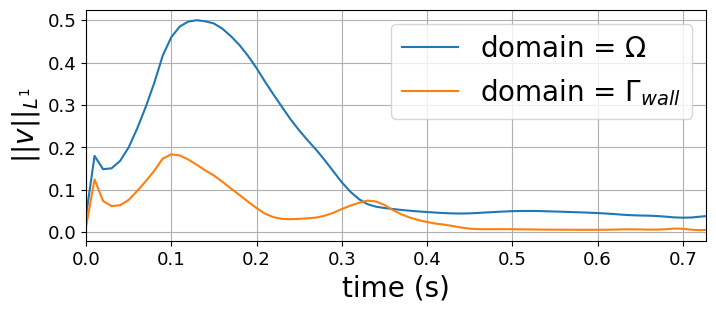} & \includegraphics[width=0.3\textwidth]{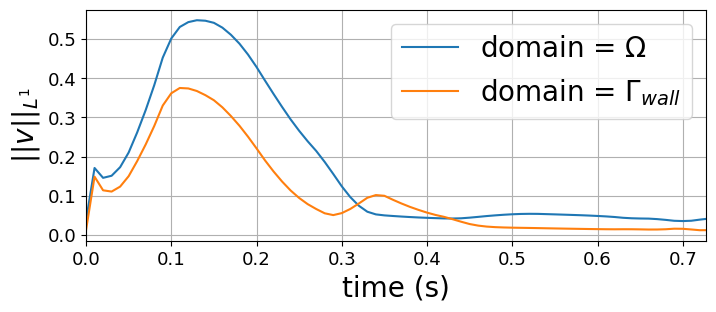} & \includegraphics[width=0.3\textwidth]{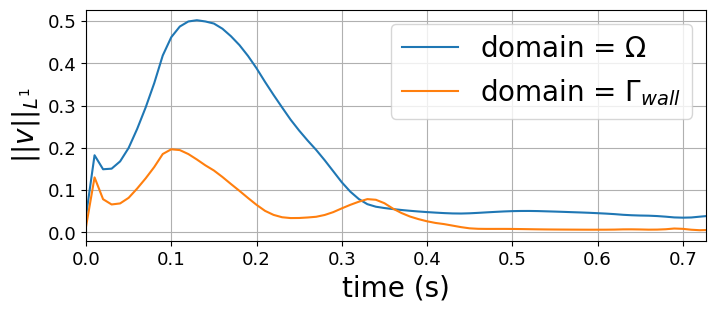}  \\
$\rnum{30.037330458411763}\%$ & $\rnum{61.34633930209336}\%$ & $\rnum{32.401548820743955}\%$ \\
\hline
\includegraphics[width=0.3\textwidth]{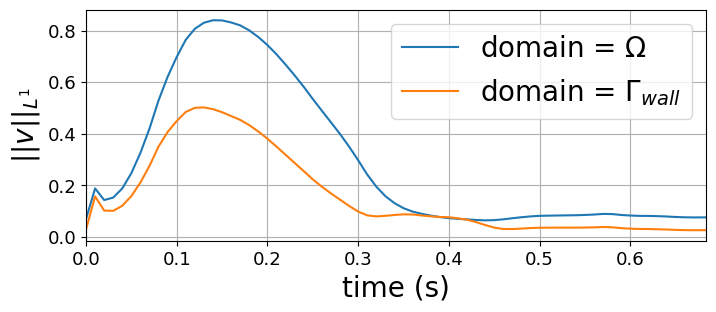} & \includegraphics[width=0.3\textwidth]{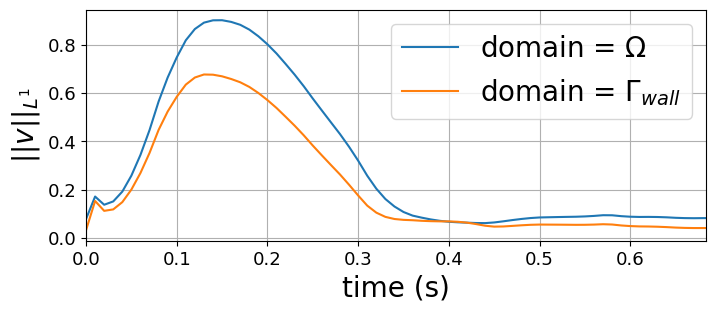} & \includegraphics[width=0.3\textwidth]{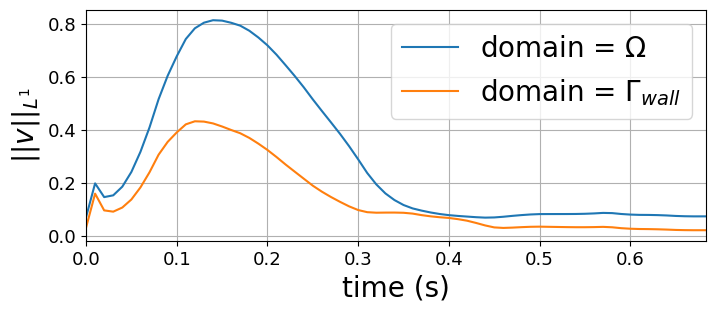}  \\
$\rnum{54.35569334548773}\%$ & $\rnum{71.14060681860734}\%$ & $\rnum{48.664951846109766}\%$ \\
\hline
\includegraphics[width=0.3\textwidth]{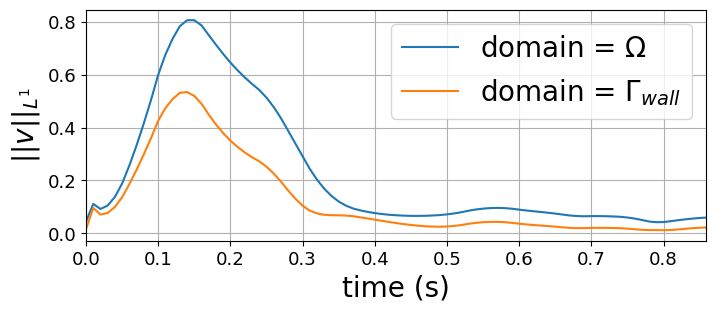} & \includegraphics[width=0.3\textwidth]{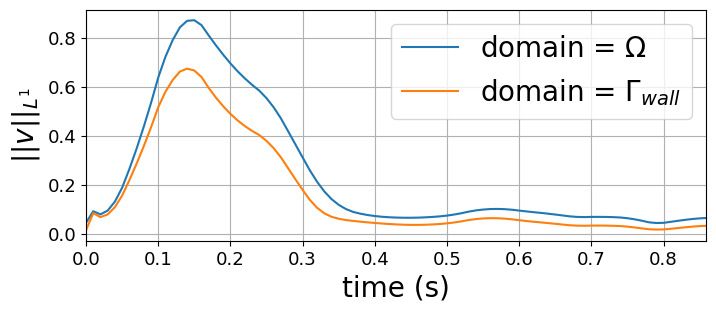} & \includegraphics[width=0.3\textwidth]{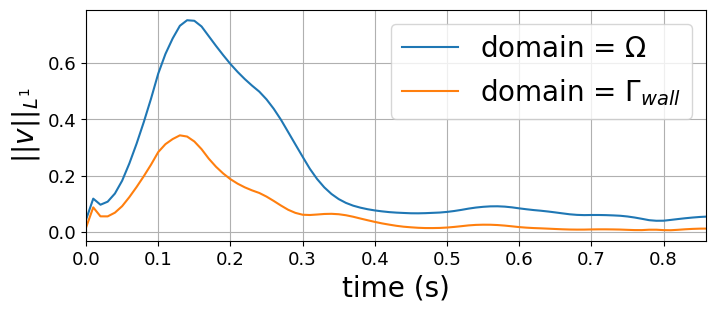}  \\
$\rnum{55.638468716537744}\%$ & $\rnum{69.57710316635993}\%$ & $\rnum{36.268238513893714}\%$ \\
\hline
\end{tabular}
    \caption{Time-dependence of the average tangential velocity over the wall domain $\Gamma_{\rm wall}$ and the average velocity over the entire domain $\Omega$ for all seven MRI datasets and three segmentation methods. Each row corresponds to one dataset. The percentages shown below the graphs represent the ratio of the average tangential velocity along the wall to the velocity averaged over the entire domain, averaged over both space and time.}
    \label{unsteady tab:v_on_wall}
\end{figure*}

Figure~\ref{unsteady tab:v_on_wall} shows the time evolution of the average tangential velocity along the vessel wall and the average velocity over the entire domain for all seven MRI datasets and three segmentation methods (manual, automatic, and automatic with one additional pixel layer). The average wall velocity $v_{\rm avg}^{\bndry{wall}}(t)$ and the domain average velocity $v_{\rm avg}^{\Omega}(t)$ are defined as
\begin{equation}
v_{\rm avg}^{\bndry{wall}}(t)=\frac{1}{|\bndry{wall}|}\int_{\bndry{wall}} |{\bf v}_{\tau}|\,{\rm d}S,\quad
v_{\rm avg}^{\Omega}(t)=\frac{1}{|\Omega|}\int_{\Omega} |{\bf v}|\,{\rm d}x.
\end{equation}
If the classical no-slip condition were valid, $v_{\rm avg}^{\bndry{wall}}(t)$ would vanish identically. Instead, the results consistently show values between 30\% and 80\% of the average domain velocity $v_{\rm avg}^{\Omega}(t)$, revealing a pronounced degree of wall slip. The percentages indicated below each graph quantify the space-time average ratio
\begin{equation}
R=\frac{\displaystyle\int_0^T v_{\rm avg}^{\bndry{wall}}(t)\,{\rm d}t}{\displaystyle\int_0^T v_{\rm avg}^{\Omega}(t)\,{\rm d}t},
\end{equation}
which provides a compact measure of slip intensity over the cardiac cycle.

\begin{figure}
    \centering
    \footnotesize
    \begin{tabular}{cccccccc}
          & D1 & D2 & D3 & D4 & D5 & D6 & D7 \\[-5mm]
         \rotatebox{90}{\parbox{3.4cm}{\centering \tiny MRI Data}} & 
         \includegraphics[width=0.11\linewidth]{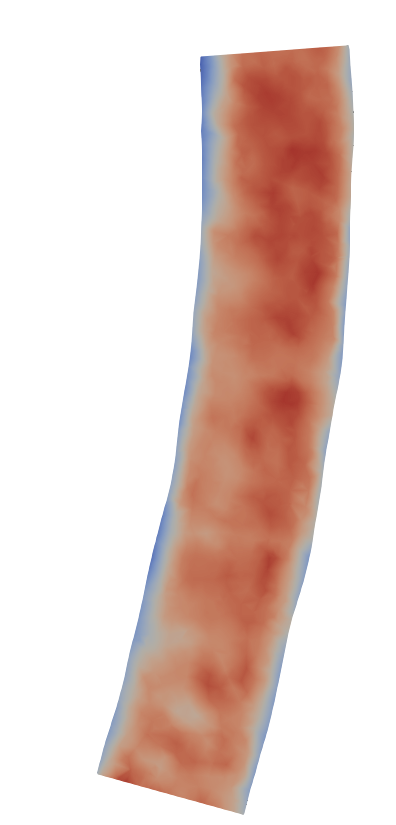} & 
         \includegraphics[width=0.11\linewidth]{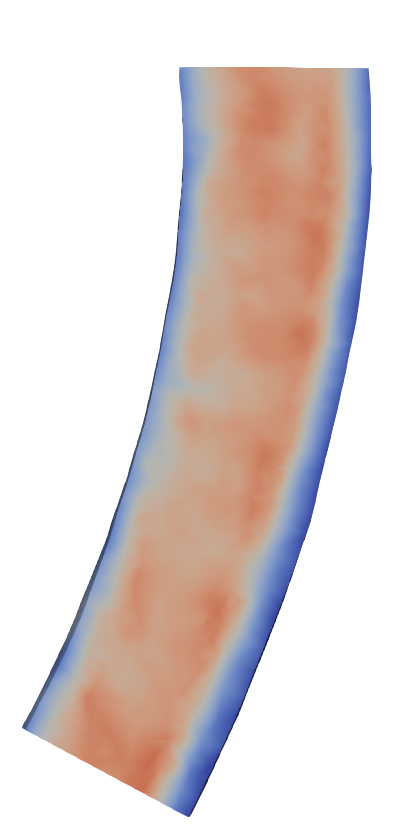} & 
         \includegraphics[width=0.11\linewidth]{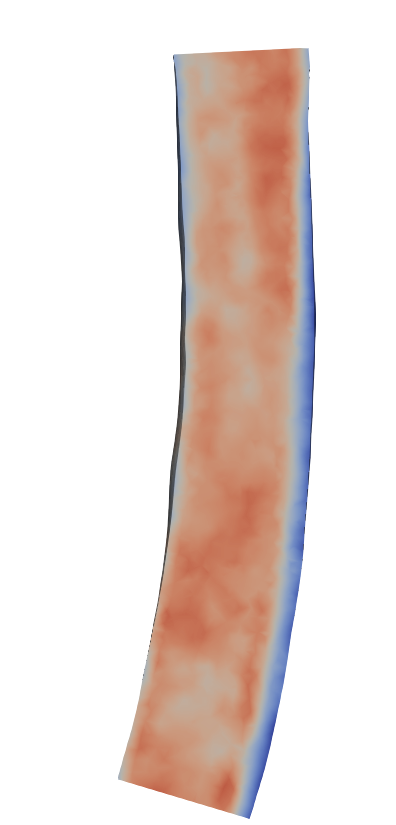} & 
         \includegraphics[width=0.11\linewidth]{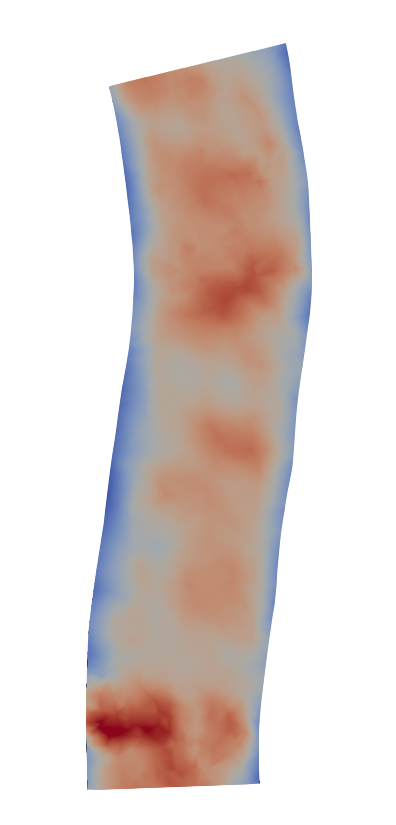} &
         \includegraphics[width=0.11\linewidth]{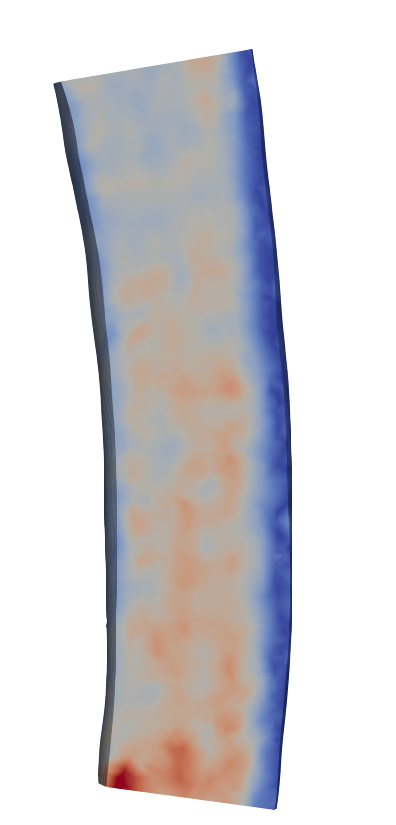} &
         \includegraphics[width=0.11\linewidth]{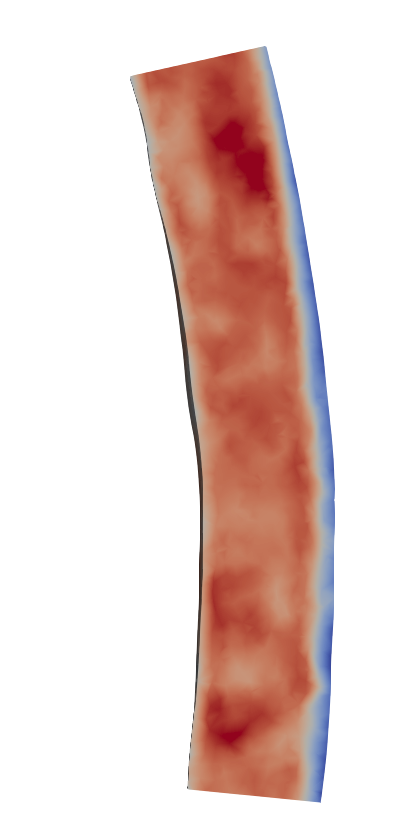} &
         \includegraphics[width=0.11\linewidth]{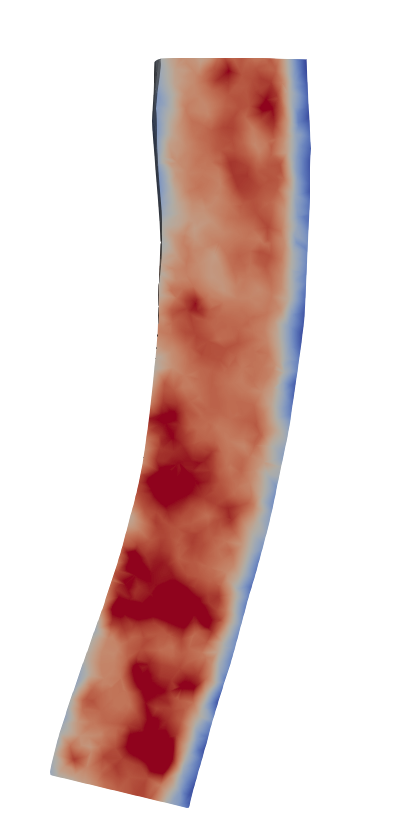}\\[-5mm]
         \rotatebox{90}{\parbox{3.4cm}{\centering  \tiny Velocity magnitude $|\mathbf{v}|$\\ no-slip}} & 
         \includegraphics[width=0.11\linewidth]{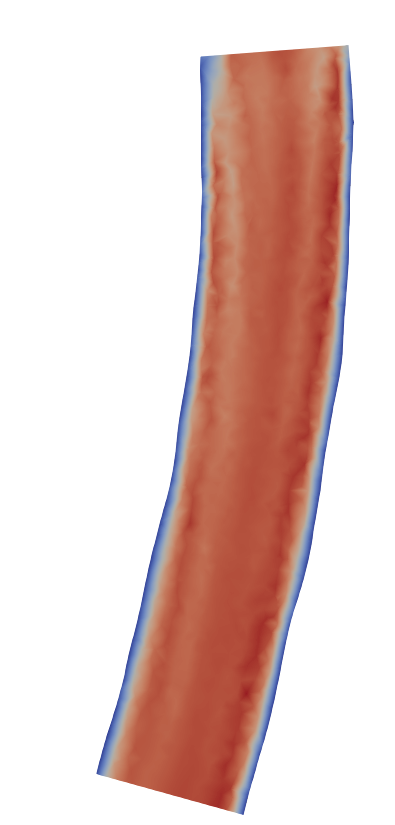} & 
         \includegraphics[width=0.11\linewidth]{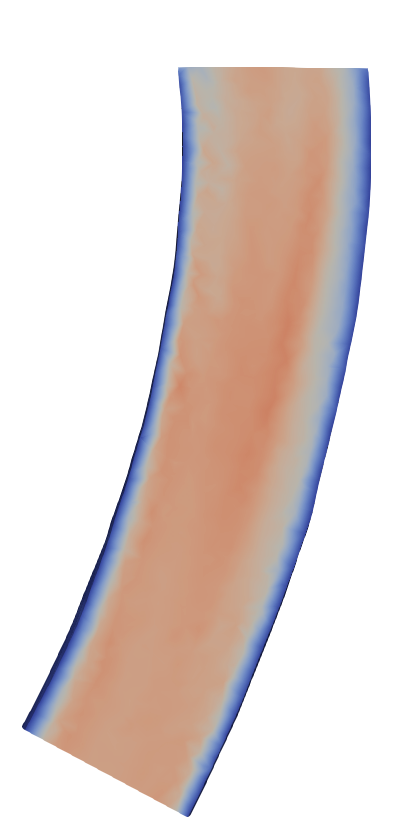} & 
         \includegraphics[width=0.11\linewidth]{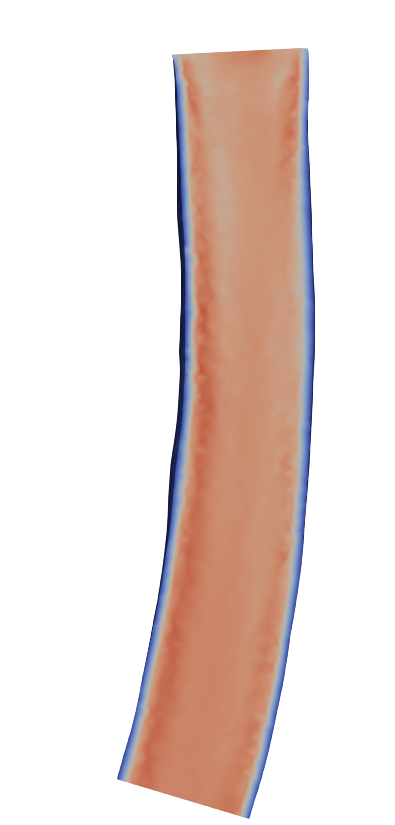} & 
         \includegraphics[width=0.11\linewidth]{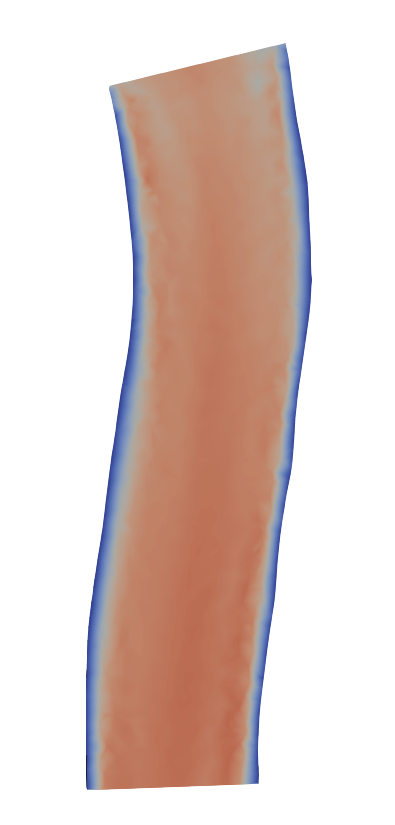} &
         \includegraphics[width=0.11\linewidth]{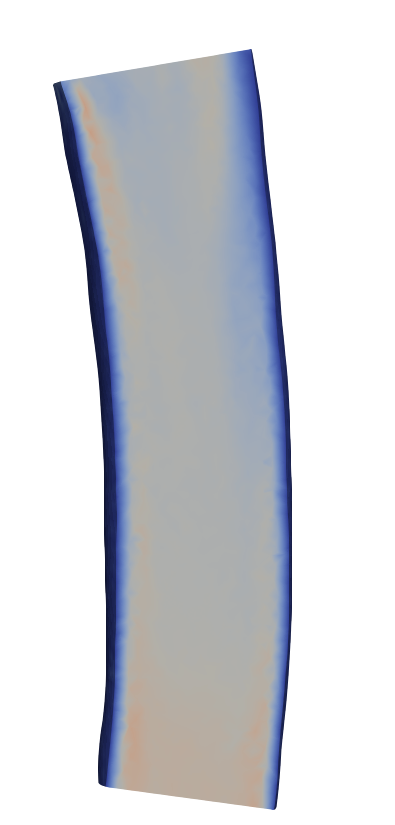} &
         \includegraphics[width=0.11\linewidth]{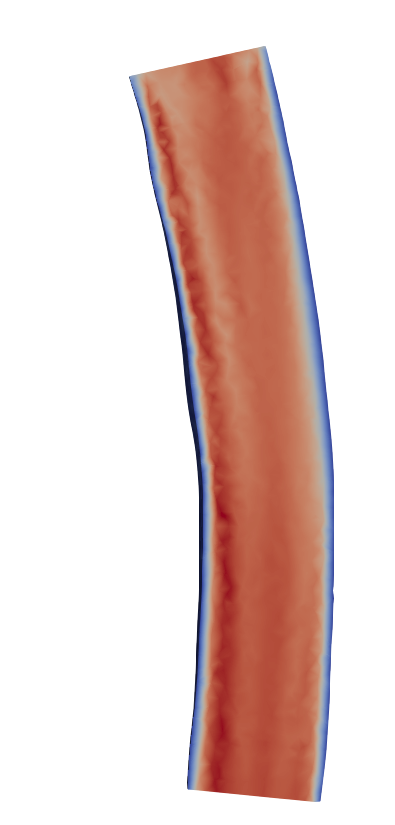} &
         \includegraphics[width=0.11\linewidth]{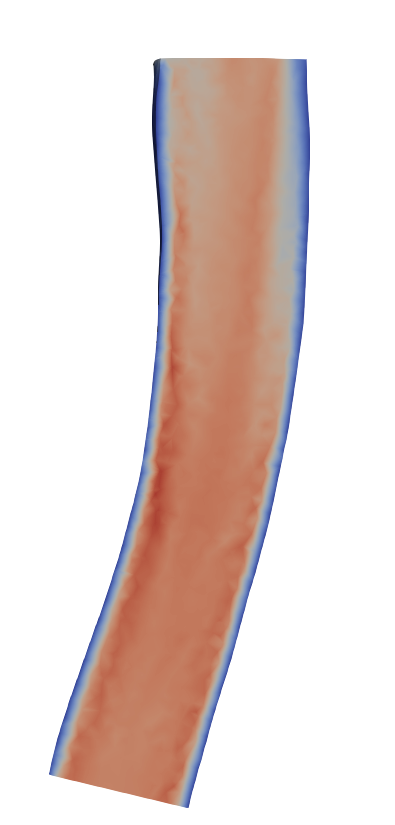}\\[-5mm]
         \rotatebox{90}{\parbox{3.4cm}{\centering  \tiny Velocity magnitude $|\mathbf{v}|$\\ constant slip}} & 
         \includegraphics[width=0.11\linewidth]{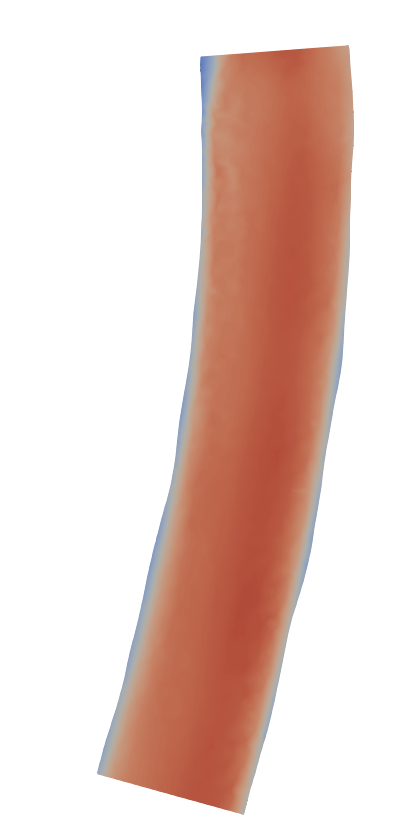} & 
         \includegraphics[width=0.11\linewidth]{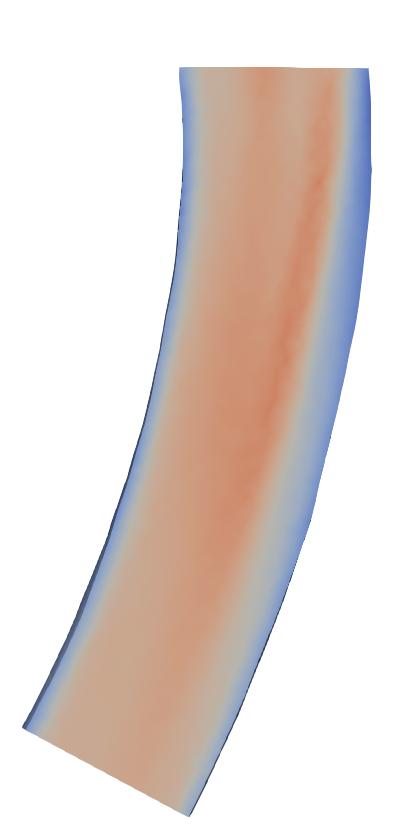} & 
         \includegraphics[width=0.11\linewidth]{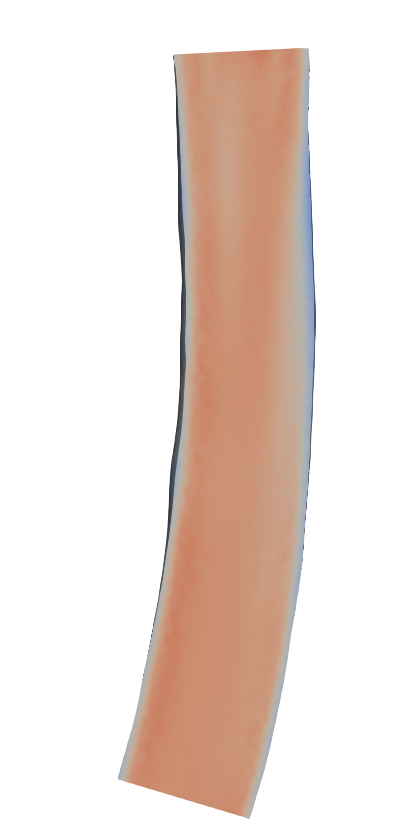} & 
         \includegraphics[width=0.11\linewidth]{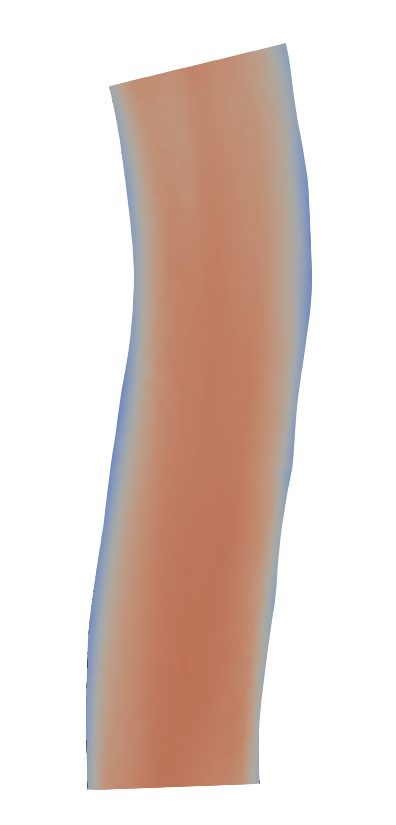} &
         \includegraphics[width=0.11\linewidth]{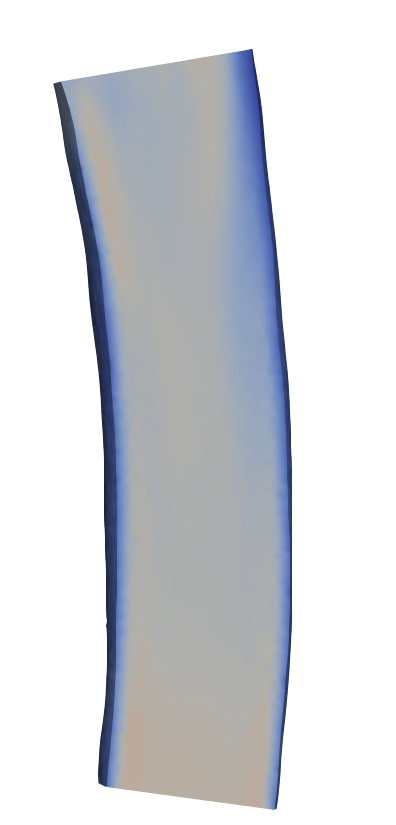} &
         \includegraphics[width=0.11\linewidth]{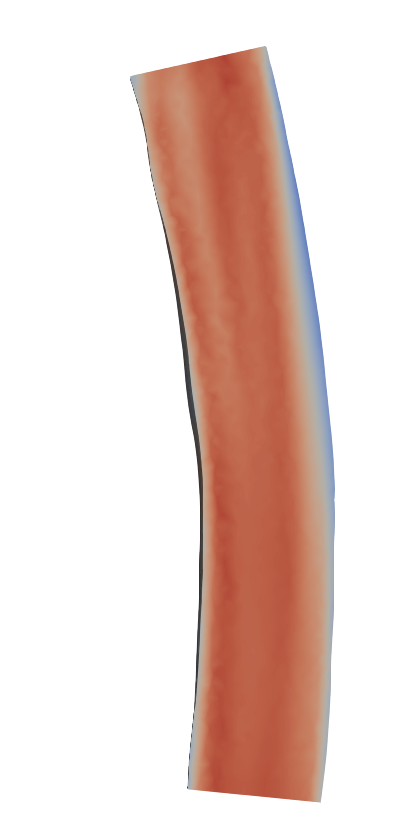} &
         \includegraphics[width=0.11\linewidth]{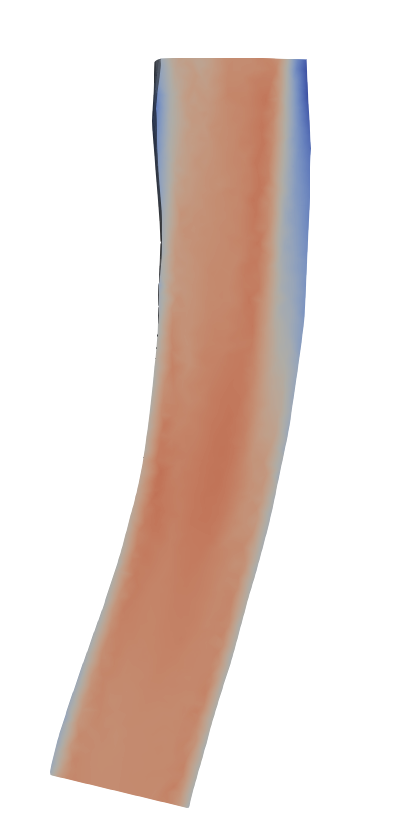}\\[-5mm]
         \rotatebox{90}{\parbox{3.4cm}{\centering  \tiny Velocity magnitude $|\mathbf{v}|$\\ spatially varying slip}} & 
         \includegraphics[width=0.11\linewidth]{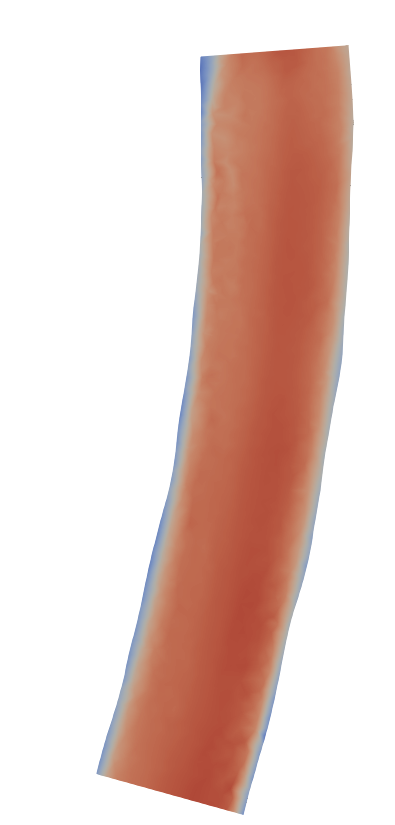} & 
         \includegraphics[width=0.11\linewidth]{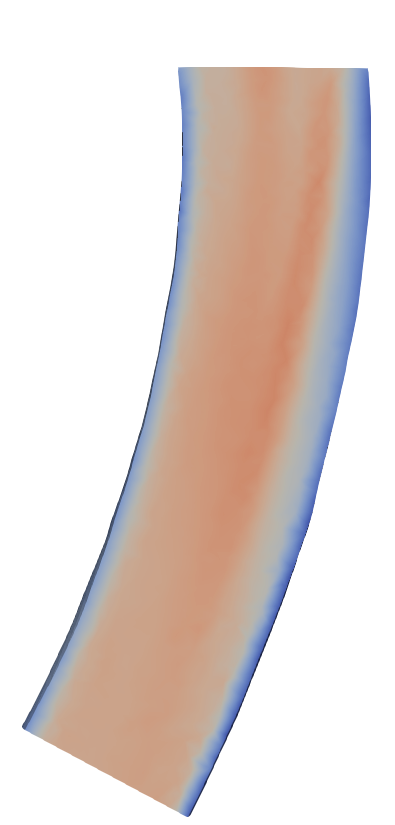} & 
         \includegraphics[width=0.11\linewidth]{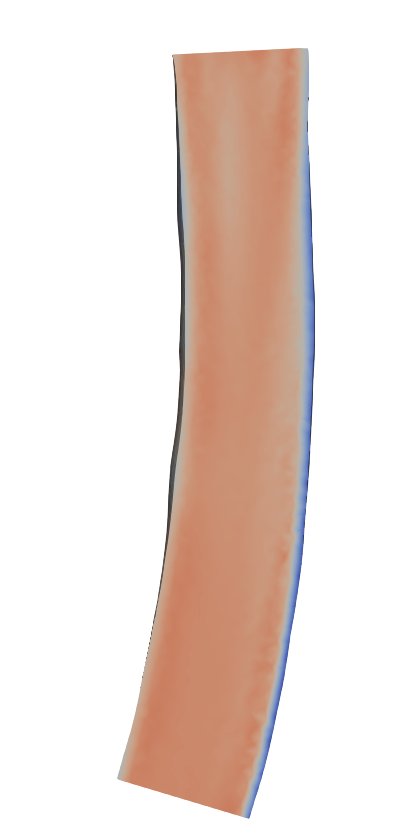} & 
         \includegraphics[width=0.11\linewidth]{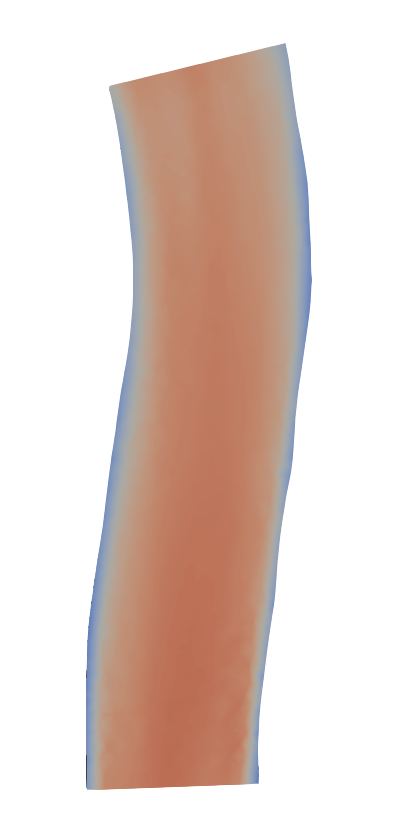} &
         \includegraphics[width=0.11\linewidth]{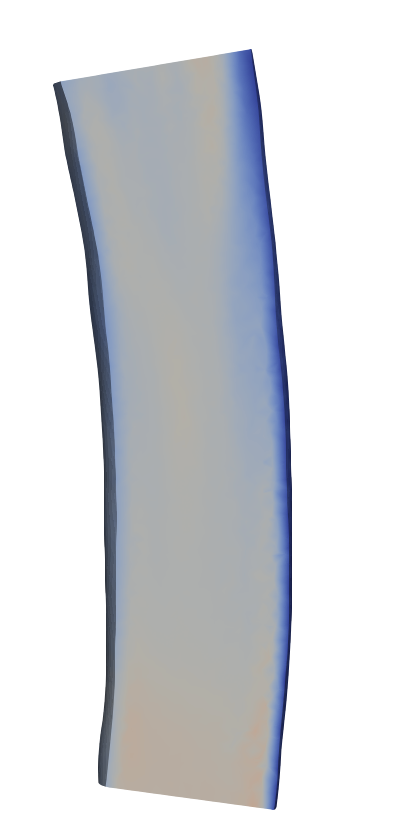} &
         \includegraphics[width=0.11\linewidth]{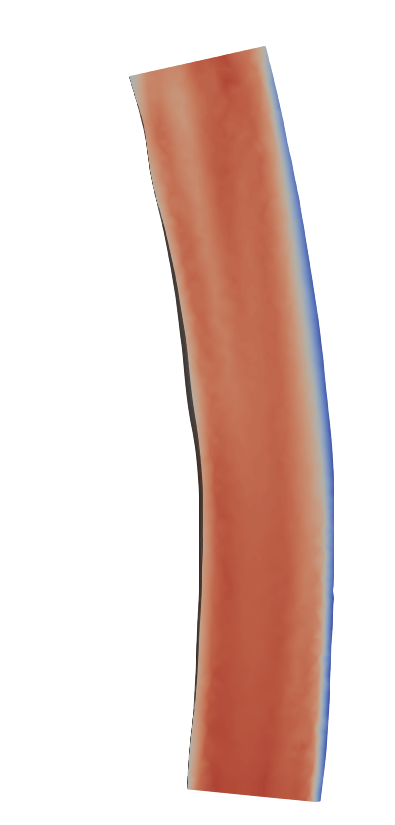} &
         \includegraphics[width=0.11\linewidth]{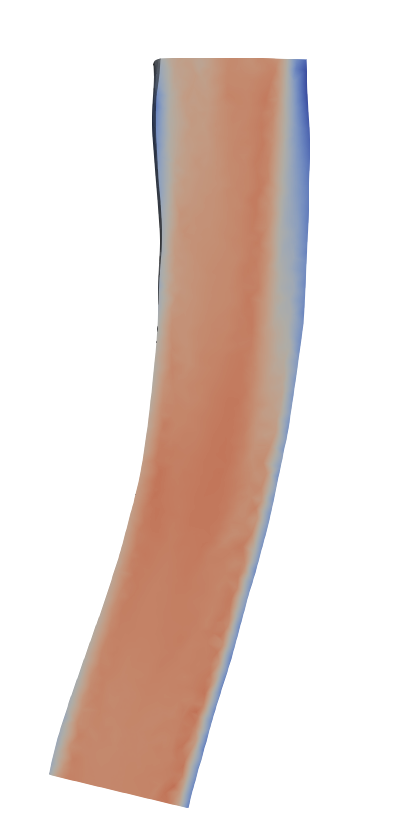}
         \end{tabular}\\
         \includegraphics[height=0.7cm]{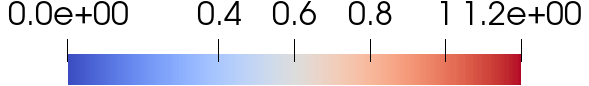}\\
         \begin{tabular}{cccccccc}
          & & & & & & & \\[-4mm]
         \rotatebox{90}{\parbox{4.0cm}{\centering  \tiny Slip parameter $\kappa_{\rm opt}(x)$\\ spatially varying slip}} & 
         \raisebox{-2.5pt}{\includegraphics[width=0.11\linewidth]{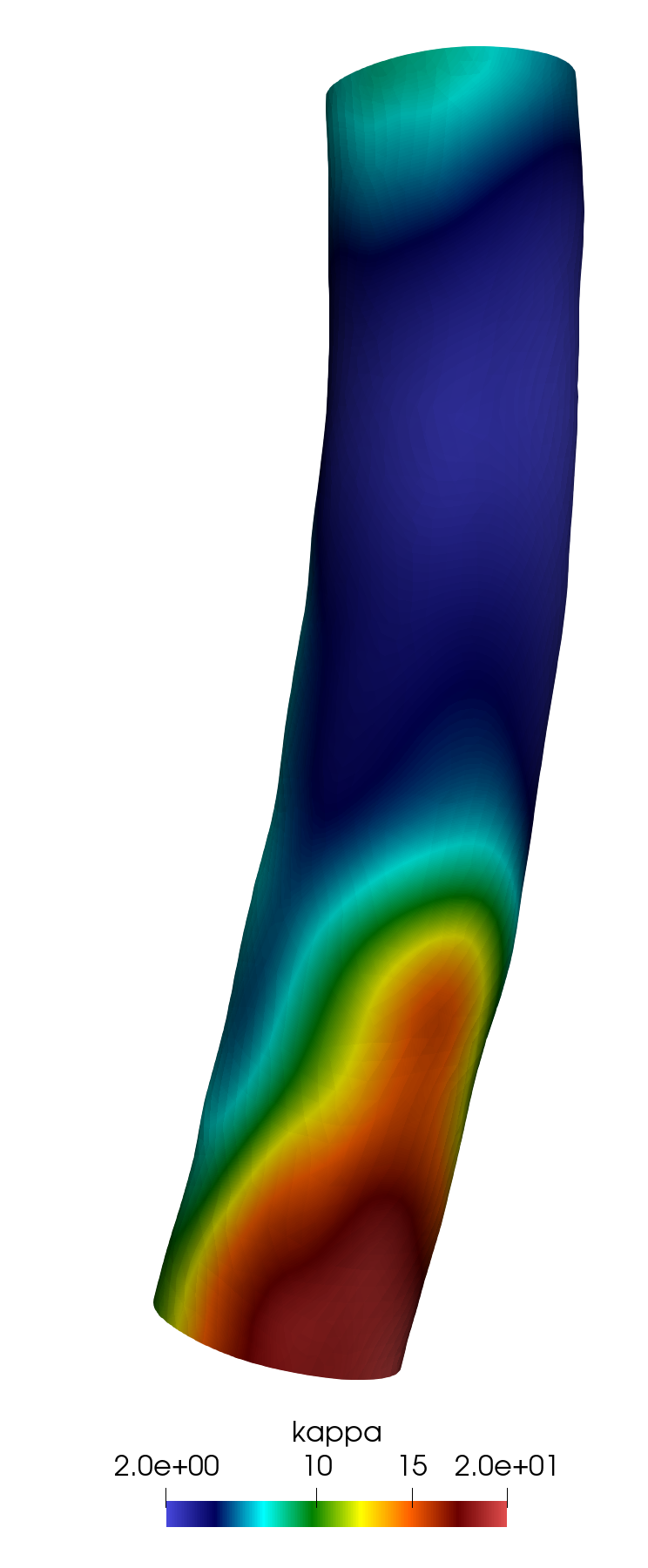}}& 
         \raisebox{-2pt}{\includegraphics[width=0.11\linewidth]{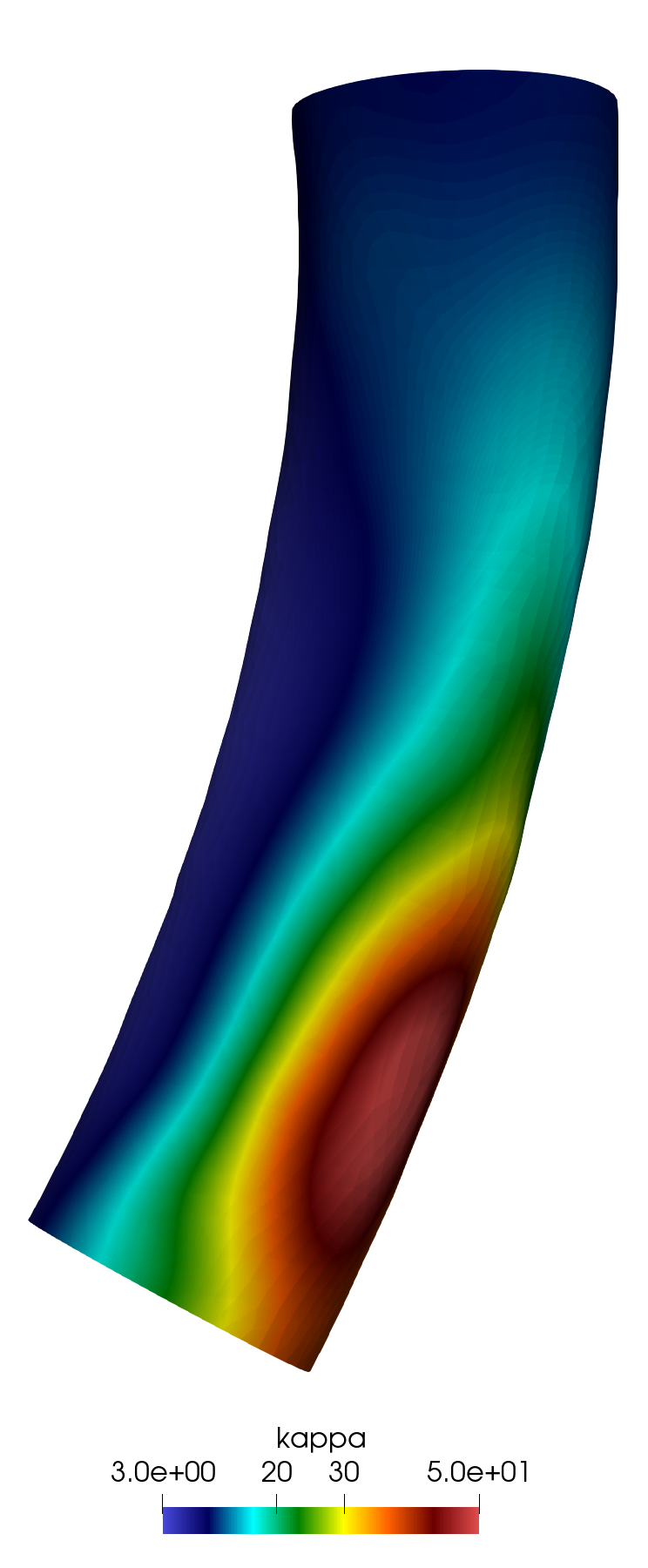}} & 
         \includegraphics[width=0.11\linewidth]{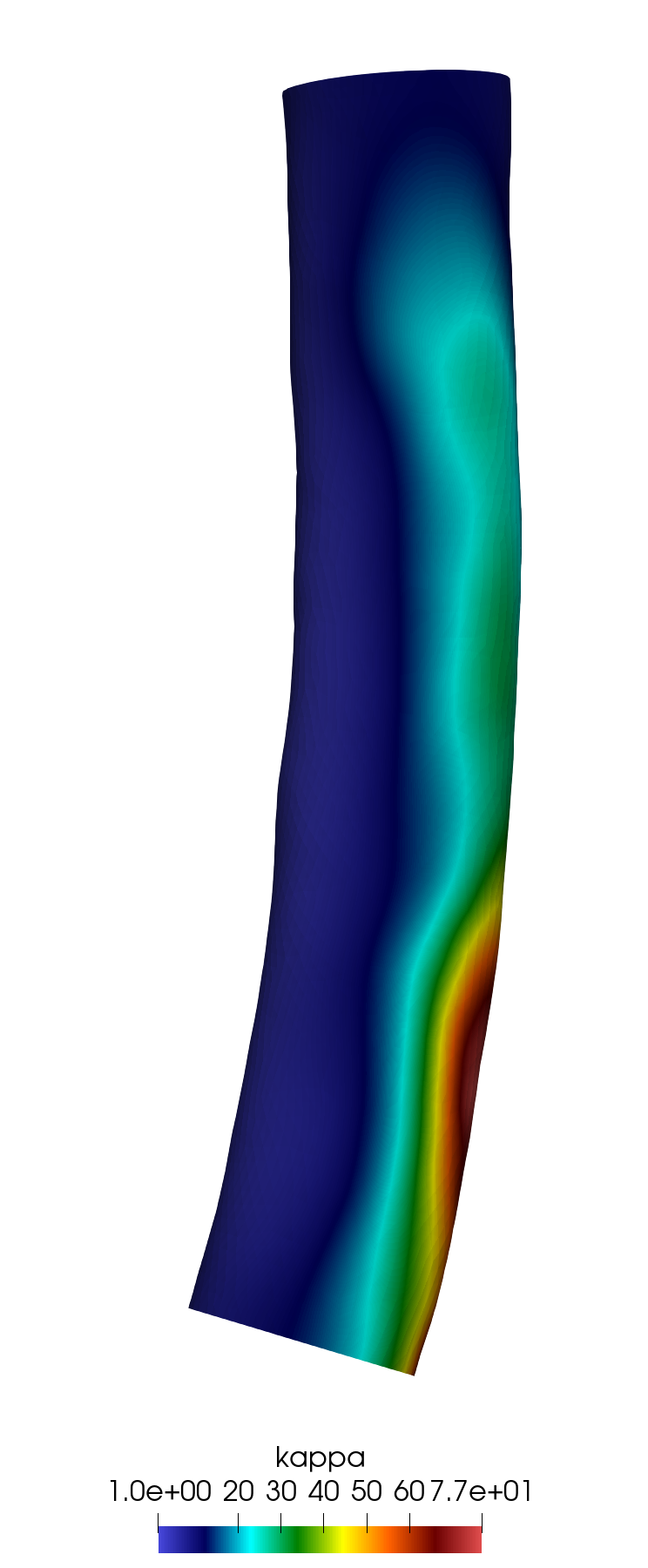} & 
         \raisebox{-5pt}{\includegraphics[width=0.11\linewidth]{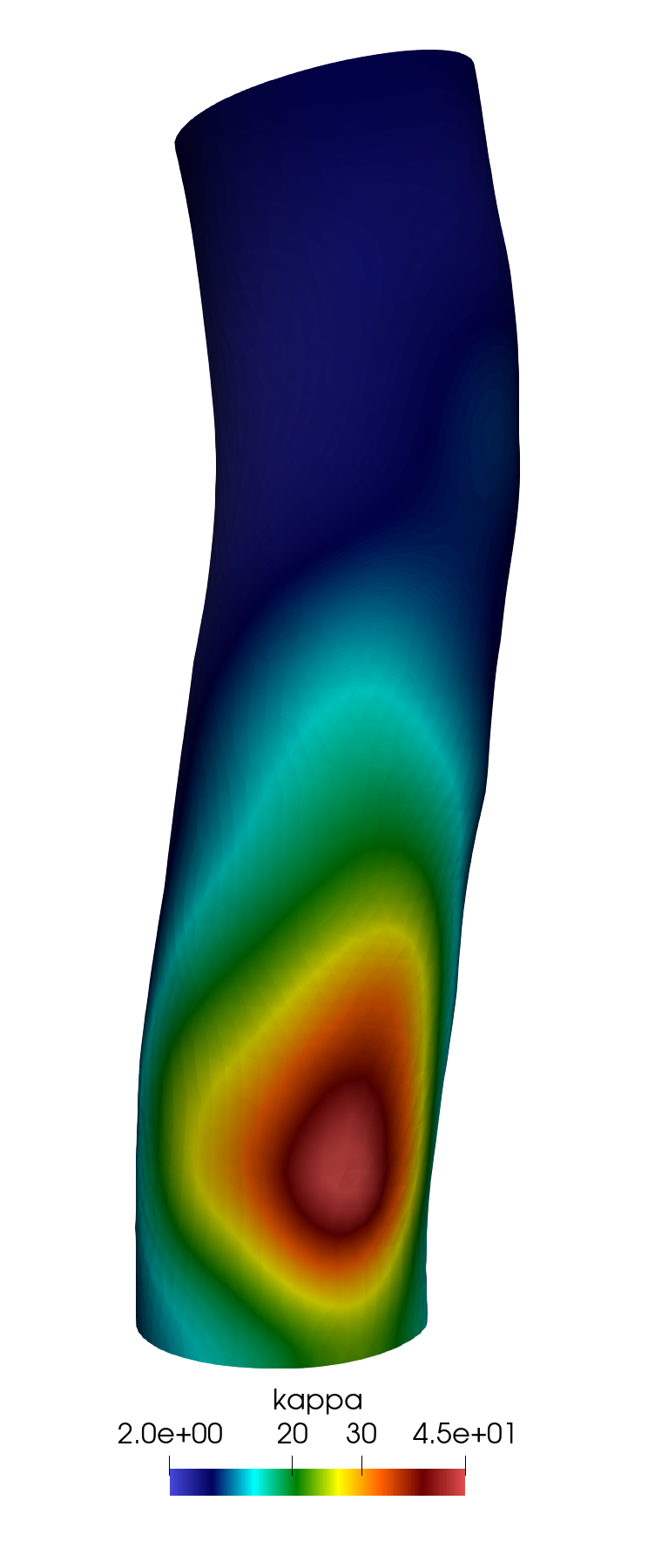}} &
         \raisebox{-4pt}{\includegraphics[width=0.11\linewidth]{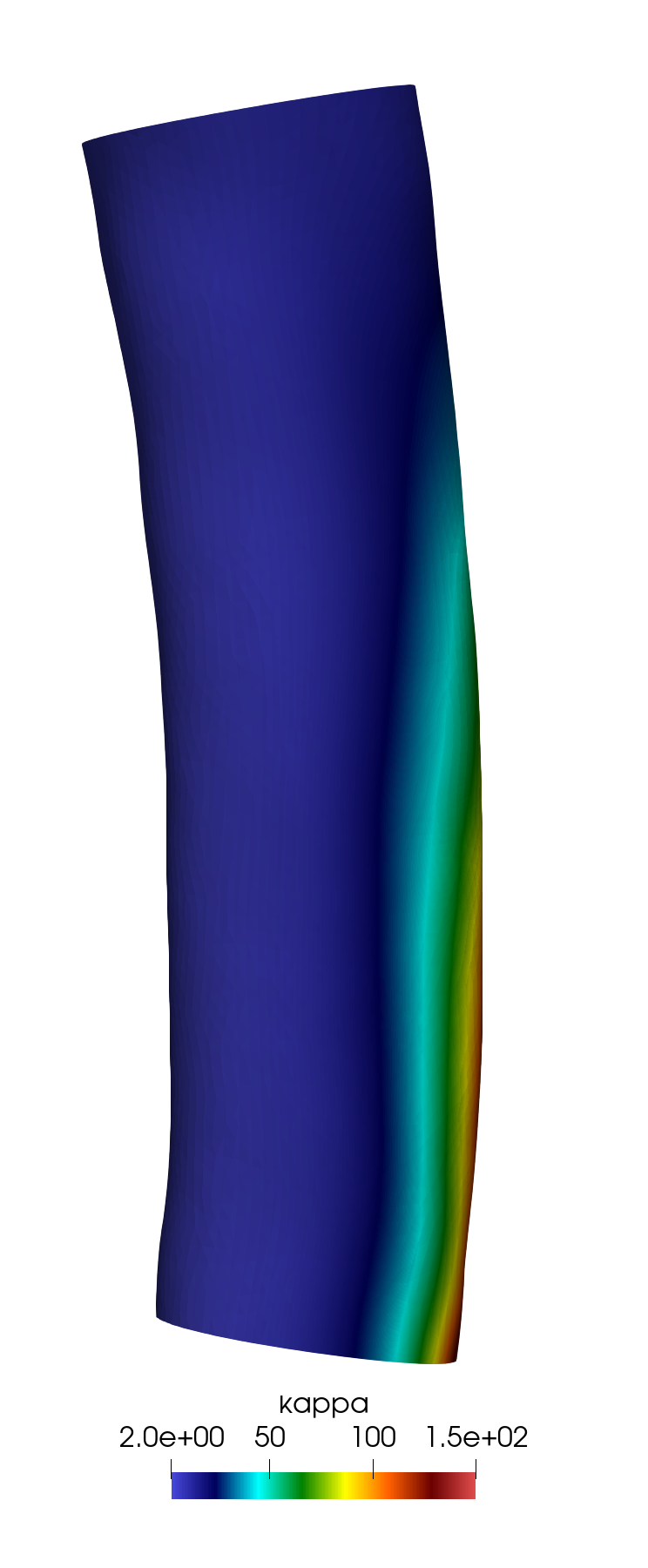}}&
         \raisebox{-0pt}{\includegraphics[width=0.11\linewidth]{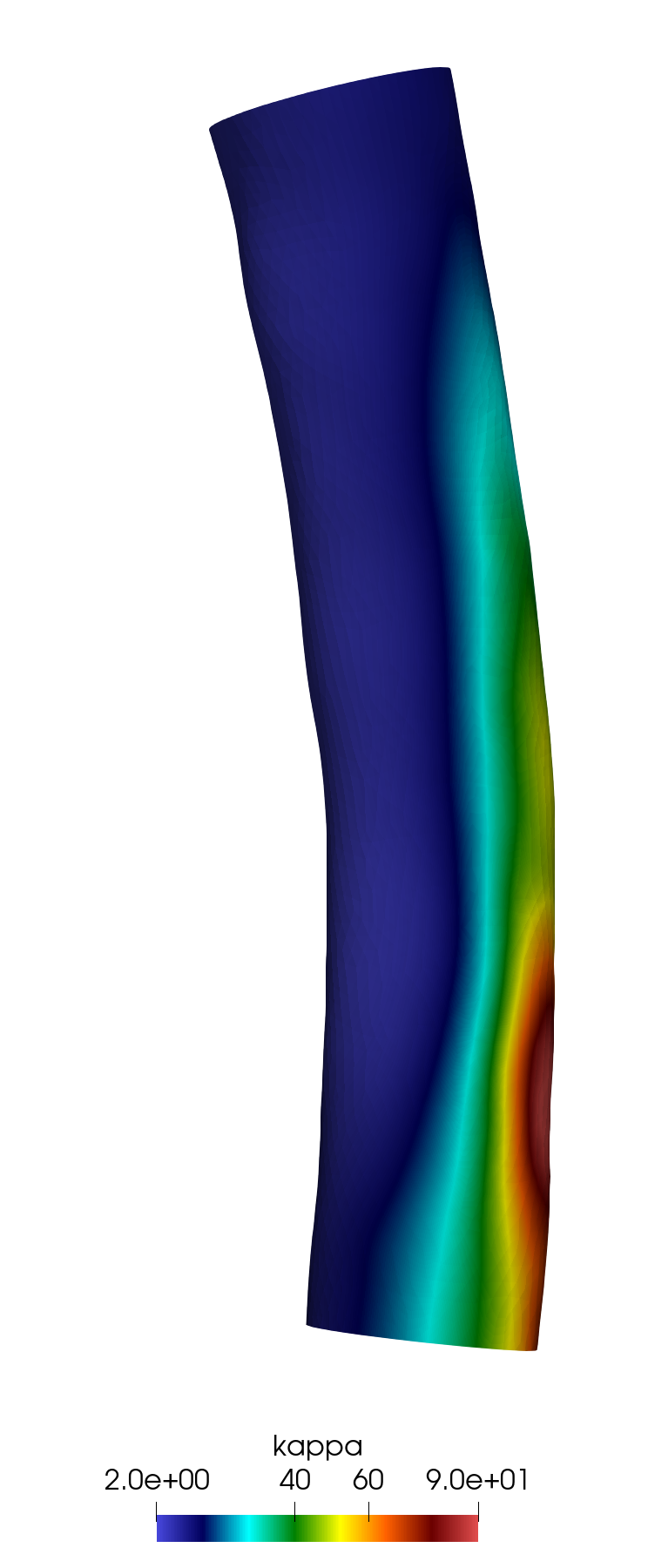}} &
         \raisebox{1pt}{\includegraphics[width=0.11\linewidth]{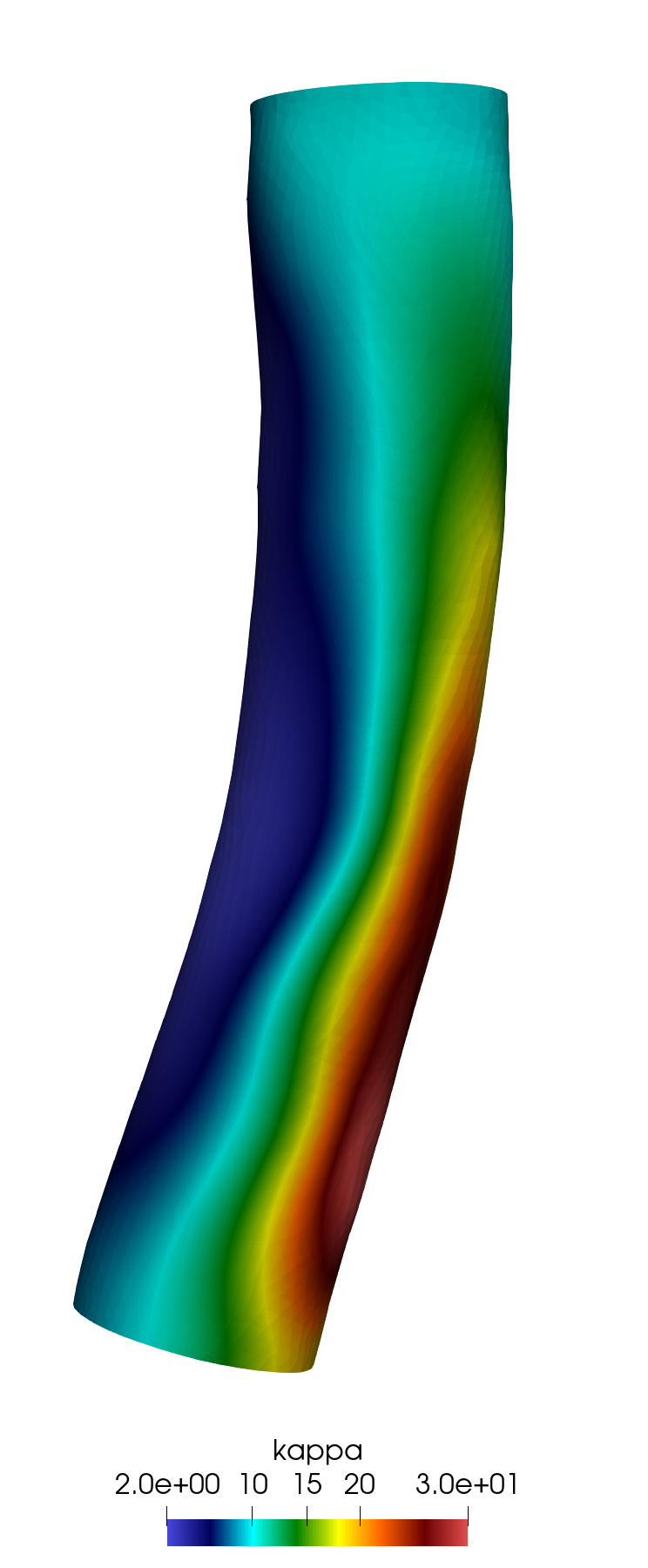}}\\
          & & & & & & & \\[-10pt]
    \end{tabular}  
        \caption{Comparison of velocity reconstructions across different boundary condition models for all datasets using the manual segmentation method. Each column corresponds to one dataset, while the rows (top to bottom) show: interpolated \ac{4DMRI} velocity data, velocity reconstruction with a no-slip boundary condition, reconstruction with a constant slip parameter $\kappa_{\rm opt}$, reconstruction with a spatially varying slip parameter $\kappa_{\rm opt}(x)$, and the corresponding spatial distribution of $\kappa_{\rm opt}(x)$.}

    \label{fig:vols results}
\end{figure}

To further investigate these findings, we apply our fitting methodology to all seven datasets and compare the reconstructed velocity fields with MRI data in \Cref{fig:vols results}. Each column corresponds to one dataset. From top to bottom, the rows show: 

\medskip

\newcommand{\Row}[2]{%
  \par\noindent\hangindent=1.4cm\hangafter=1
  {Row #1:} #2\par
}

\noindent 
\Row{1}{Velocity fields interpolated from MRI data.}
\Row{2}{Velocity reconstructions with the classical no-slip condition (in which only $\mathbf{v}_{\rm in}$ is fitted via the above method).}
\Row{3}{Velocity reconstructions with a constant optimally fitted slip parameter $\kappa_{\rm opt}$ (both $\mathbf{v}_{\rm in}$ and $\kappa_{\rm opt}$ are fitted).}
\Row{4}{Velocity reconstructions with a spatially varying slip parameter $\kappa_{\rm opt}(x)$ ($\mathbf{v}_{\rm in}$ and spatially varying $\kappa_{\rm opt}(x)$ are fitted).}
\Row{5}{Spatial distributions of $\kappa_{\rm opt}(x)$ corresponding to the results presented in Row 4.}

\medskip

It appears that the error functional $\mathcal{J}$ is significantly lower when the slip parameter is optimally fitted (Row 3) compared to the prescribed no-slip condition (Row 2). Except for Dataset 5, where the blood slip is minimal, the values of $\mathcal{J}$ for the optimal slip case range between 60\% and 80\% of those obtained under the no-slip condition, indicating that allowing some degree of slip provides a markedly better description of the observed \emph{in vivo} data, see Table~\ref{tab:costJ}.

The velocity reconstructions obtained with constant (shown in a Row 3 of Figure~\ref{fig:vols results}) and spatially varying slip (shown in a Row 4 of Figure~\ref{fig:vols results}) models are highly similar, yet the slip distributions differ considerably: over most of the wall, $\kappa$ remains close to zero, indicating nearly perfect slip, while in localized regions $\kappa$ grows very large, effectively enforcing a no-slip condition. This demonstrates that only a small fraction of the boundary substantially resists the flow.

The optimal slip parameters $\kappa_{\rm opt}$ for each dataset and segmentation method are summarized in \Cref{tab:pat_kappa}, the corresponding slip lengths $\beta = \mu / \kappa_{\rm opt}$ in \Cref{tab:pat_beta}, and the average tangential velocities during systole in \Cref{tab:pat_vavg}. For $\kappa = 200$, the corresponding slip length is on the order of the erythrocyte diameter, below which blood cannot reasonably be modeled as a continuum. Thus, $\kappa = 200$ effectively represents a no-slip condition in the context of continuum model, while values $\kappa < 20$ clearly indicate that blood does not adhere to the vessel wall. Across all datasets and segmentation methods, the fitted values of $\kappa$ remain below 19 (five of seven datasets have $\kappa$ less than ten for all three segmentation methods), and the tangential wall velocities are well above zero, confirming that the no-slip condition is not valid \emph{in vivo}. Furthermore, the automatic segmentation yields systematically lower values of $\kappa$ than the manual segmentation, corresponding to higher tangential wall velocities.

\vspace*{-5pt}
\vspace*{-10pt}
\begin{table}
    \centering
    \caption{Fitted slip parameters $\kappa_{\rm opt}$ [\unit{Pa.s.m^{-1}}] for different MRI datasets and different segmentation methods.}
    \label{tab:pat_kappa}
\vspace*{-5pt} 
\begin{ruledtabular}
\begin{tabular}{p{6cm}lllllll}
    Dataset & D1 & D2 & D3 & D4 & D5 & D6 & D7\\ 
    \midrule
    Manual segmentation &
    3.89 & 5.42 & 5.29 & 4.96 & 18.26 & 5.35 & 4.59 \\
    Automatic segmentation &
    1.62 & 3.09 & 2.14 & 1.49 & 4.40 & 2.49 & 2.22\\
    Automatic segmentation with an additional layer &
    4.60 & 8.91 & 5.81 & 4.28 & 16.29 & 7.93 & 13.32 \\
    \end{tabular}
\end{ruledtabular}
\end{table}
\vspace*{-5pt}
\begin{table}
\centering
\caption{Slip length $\beta_{\rm opt} = \mu / \kappa_{\rm opt}$ [\unit{mm}] corresponding to fitted slip parameters $\kappa_{\rm opt}$.}\label{tab:pat_beta}
\vspace*{-5pt}
\begin{ruledtabular}
\begin{tabular}{p{6cm}lllllll} 
Dataset & D1 & D2 & D3 & D4 & D5 & D6 & D7 \\ 
\midrule
Manual segmentation & 1.0 & 0.72 & 0.74 & 0.79 & 0.21 & 0.73 & 0.85 \\ 
Automatic segmentation & 2.41 & 1.26 & 1.82 & 2.61 & 0.89 & 1.56 & 1.76 \\ 
Automatic segmentation with an additional layer & 0.85 & 0.44 & 0.67 & 0.91 & 0.24 & 0.49 & 0.29 \\
\end{tabular}
\end{ruledtabular}
\end{table}

\begin{table}
\centering
\caption{Fitted average tangential velocities in systole $v_{\rm avg, sys}^{\bndry{wall}}$ [\unit{m/s}] for different MRI datasets and different segmentation methods. The methods include manual segmentation, automatic segmentation using the TotalSegmentator tool, and automatic segmentation with an expanded domain by adding one additional layer of pixels (1 px = 1.05 mm in the MRI data).}
\label{tab:pat_vavg}
\begin{ruledtabular}
\begin{tabular}{p{6cm}lllllll} 
Dataset & D1 & D2 & D3 & D4 & D5 & D6 & D7 \\ 
\midrule
Manual segmentation & 0.53 & 0.36 & 0.48 & 0.45 & 0.16 & 0.50 & 0.52 \\ 
Automatic segmentation & 0.74 & 0.51 & 0.66 & 0.63 & 0.37 & 0.67 & 0.67 \\ 
Automatic segmentation with an additional layer & 0.53 & 0.30 & 0.46 & 0.47 & 0.17 & 0.42 & 0.34 \\
\end{tabular}
\end{ruledtabular}
\end{table}

\begin{table}
\centering
\caption{Value of the functional $\mathcal{J}$.}\label{tab:costJ}
\begin{ruledtabular}
\begin{tabular}{p{4cm}lllllll} 
Dataset & D1 & D2 & D3 & D4 & D5 & D6 & D7 \\ 
\midrule
Optimal slip & 
\num{0.08753098372327343} &
\num{0.0514052047293505}  &
\num{0.11238567384462418} &
\num{0.07926507793269534} &
\num{0.12609106223752242} &
\num{0.10696148503504008} &
\num{0.08309386523471203} \\
No-slip $0.98$ &
\num{0.12746338798677223} &
\num{0.06440583520878364} &
\num{0.15295115009164503} &
\num{0.08959527601701534} &
\num{0.12813611863189367} &
\num{0.14552150904923825} &
\num{0.11736148639784558}\\
Optimal/noslip & 68.6\% & 79.8\% & 73.5\% & 79.6\% & 98.4\% & 73.8\% & 70.8\%\\
\end{tabular}
\end{ruledtabular}
\end{table}

\subsection{Implications of no-slip failure: calculation of wall shear stress}
We further quantify the wall shear stress (WSS) and its average defined as
\begin{equation}
{\rm WSS}(t,x)=|(\mathbb{T}{\bf n})_{\tau}|,\qquad 
{\rm WSS}_{\rm avg}(t)=\frac{1}{|\Gamma_{\rm wall}|}\int_{\Gamma_{\rm wall}}{\rm WSS}(t,x)\,{\rm d}S,
\end{equation}
where $(\mathbb{T}{\bf n})_\tau$ is the outcome of our data assimilation method \eqref{Eq:DA_method}. 
We compare the WSS computed using the optimally fitted slip parameter $\kappa_{\rm opt}$ with that obtained under the conventional no-slip assumption.

\begin{figure}
    \centering
    \footnotesize
    \begin{tabular}{cccc}
          & Graph of average WSS on time & WSS with optimal slip [Pa]& WSS with no-slip [Pa]\\[0pt]
          & & (range 0 -- 3 Pa) & (range 0 -- 6 Pa)\\[0pt]
          & & \includegraphics[width=0.1\linewidth]{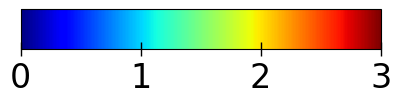} & \includegraphics[width=0.1\linewidth]{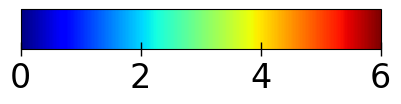} \\[-10pt]
         \rotatebox{90}{\parbox{3.0cm}{\centering Dataset 1}} & 
         \includegraphics[width=0.2\linewidth]{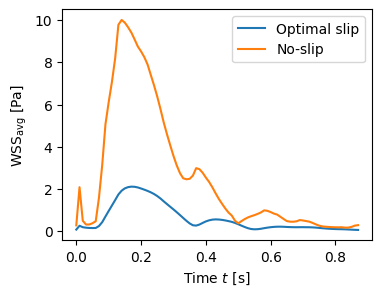} & 
         \includegraphics[height=0.1\textheight]{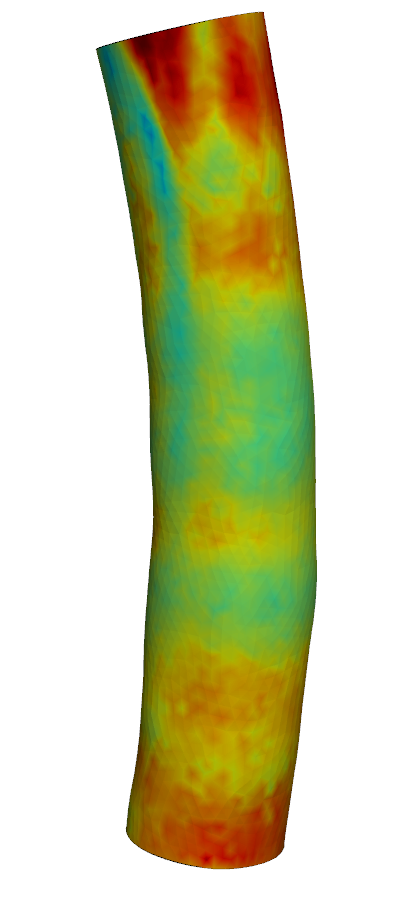} & 
         \includegraphics[height=0.1\textheight]{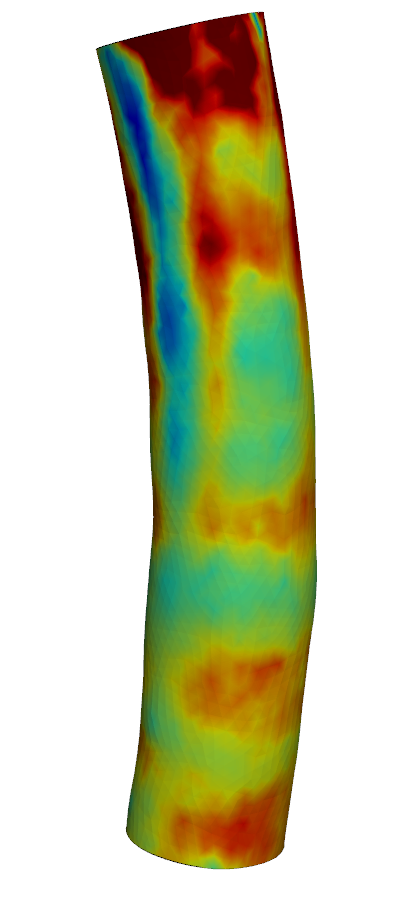} \\[-10pt]
         \rotatebox{90}{\parbox{3.0cm}{\centering Dataset 2}} & 
         \includegraphics[width=0.2\linewidth]{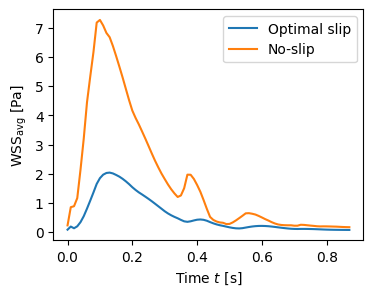} & 
         \includegraphics[height=0.1\textheight]{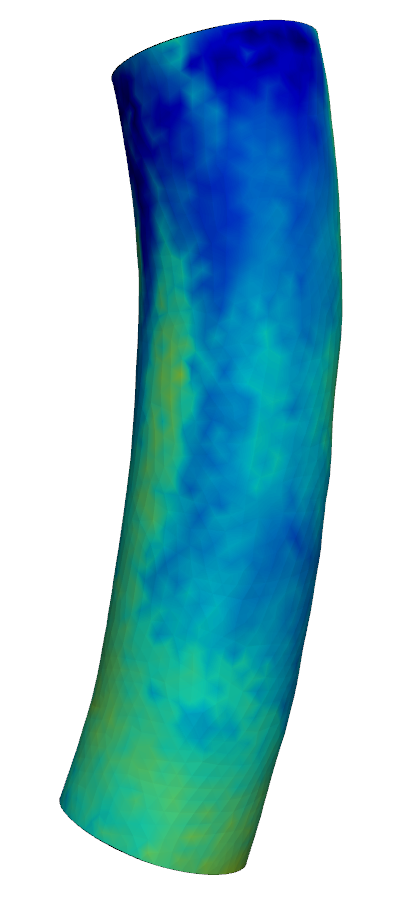} & 
         \includegraphics[height=0.1\textheight]{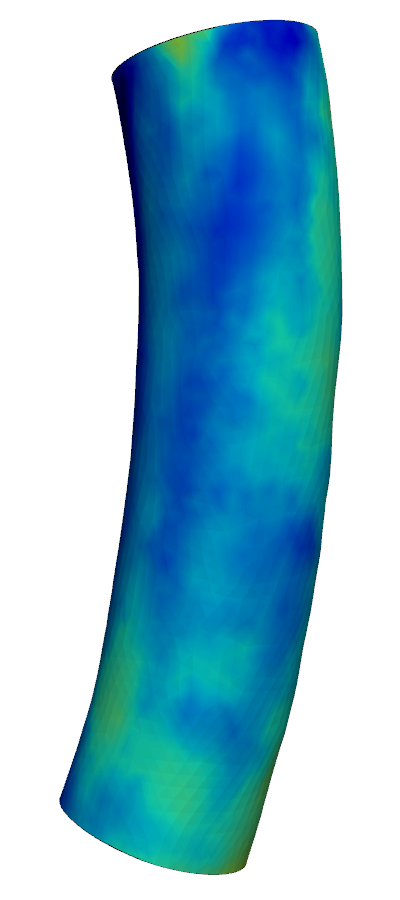} \\[-10pt]
         \rotatebox{90}{\parbox{3.0cm}{\centering Dataset 3}} & 
         \includegraphics[width=0.2\linewidth]{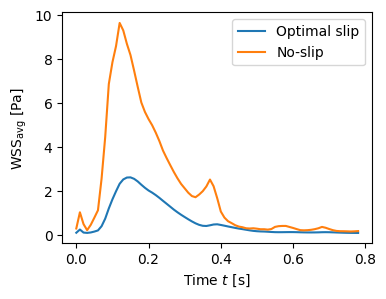} & 
         \includegraphics[height=0.1\textheight]{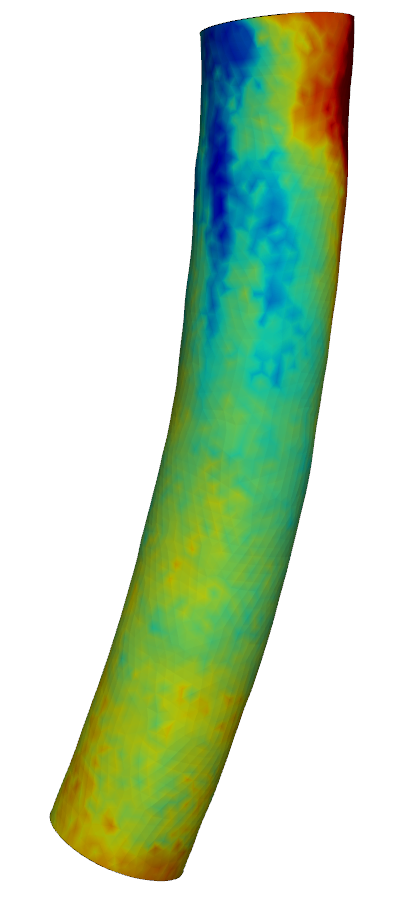} & 
         \includegraphics[height=0.1\textheight]{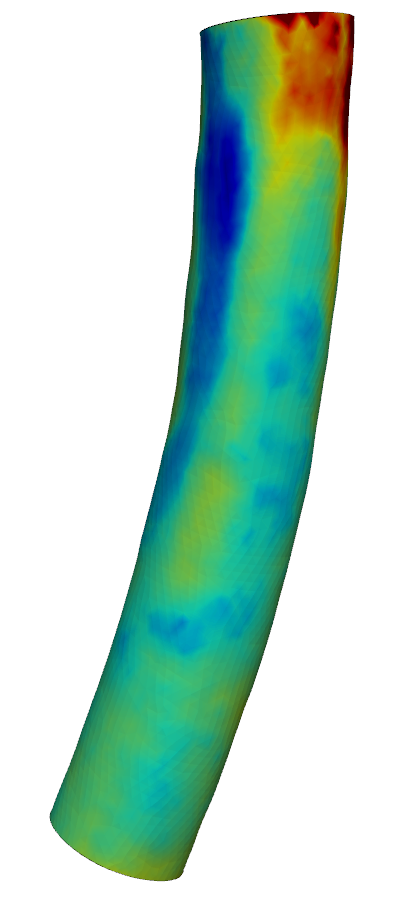} \\[-10pt]
         \rotatebox{90}{\parbox{3.0cm}{\centering Dataset 4}} & 
         \includegraphics[width=0.2\linewidth]{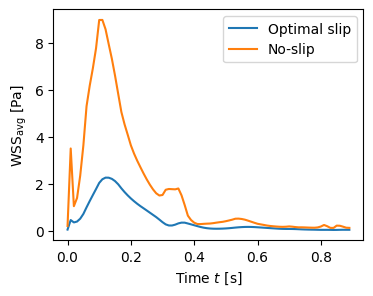} & 
         \includegraphics[height=0.1\textheight]{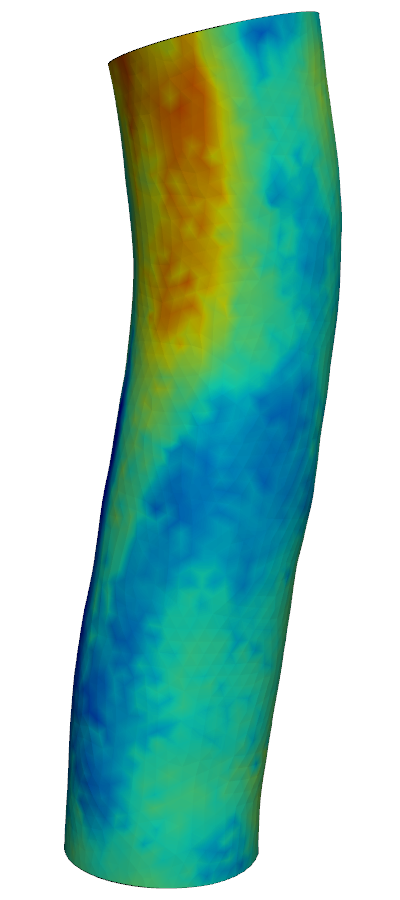} & 
         \includegraphics[height=0.1\textheight]{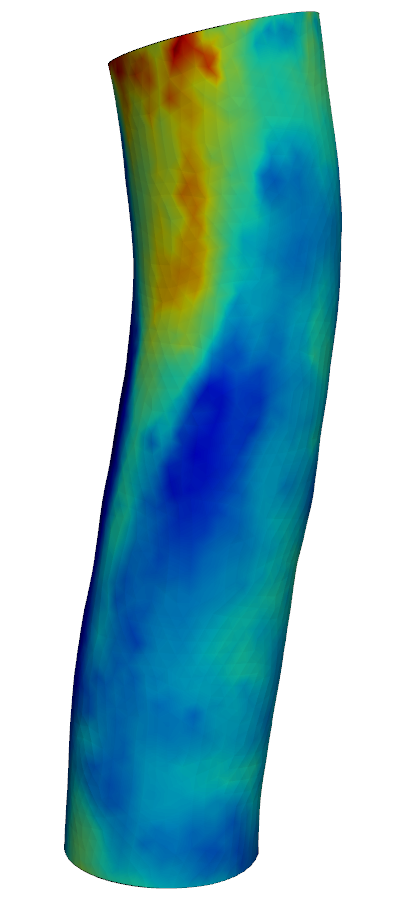} \\[-10pt]
         \rotatebox{90}{\parbox{3.0cm}{\centering Dataset 5}} & 
         \includegraphics[width=0.2\linewidth]{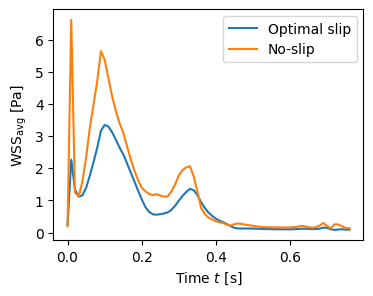} & 
         \includegraphics[height=0.1\textheight]{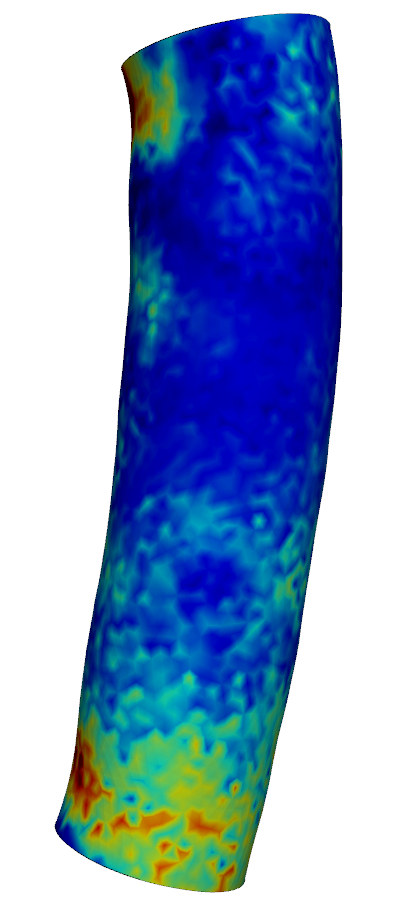} & 
         \includegraphics[height=0.1\textheight]{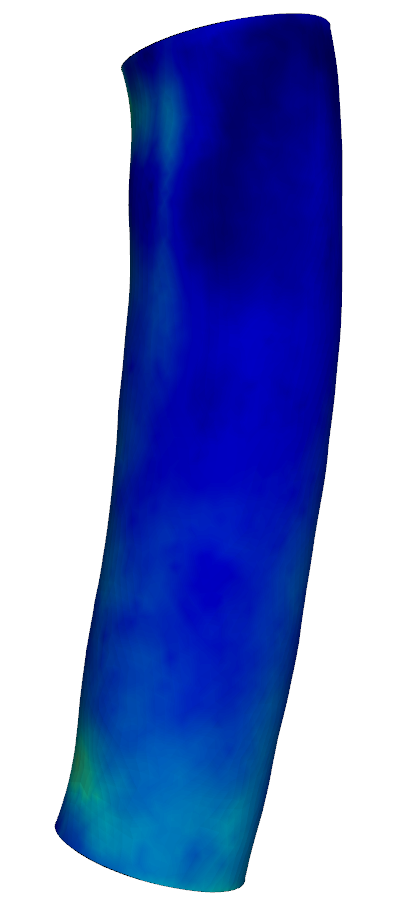} \\[-10pt]
         \rotatebox{90}{\parbox{3.0cm}{\centering Dataset 6}} & 
         \includegraphics[width=0.2\linewidth]{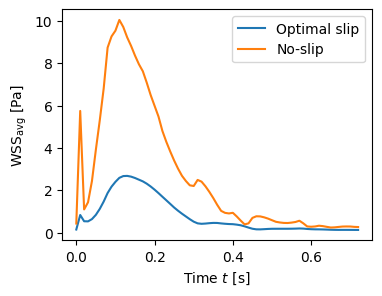} & 
         \includegraphics[height=0.1\textheight]{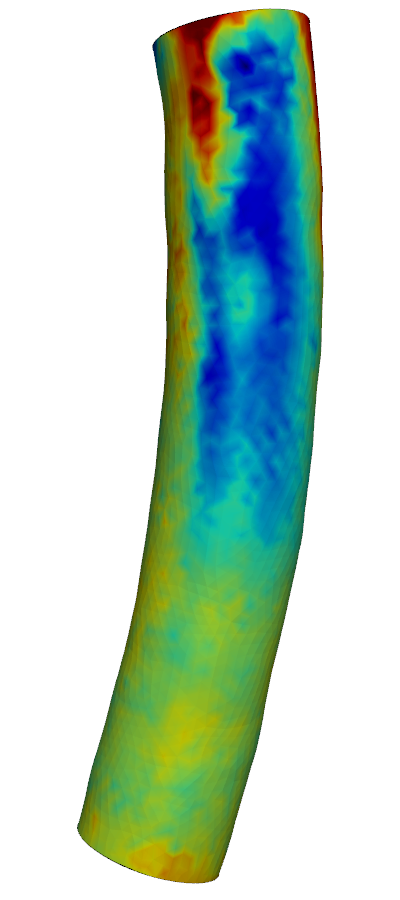} & 
         \includegraphics[height=0.1\textheight]{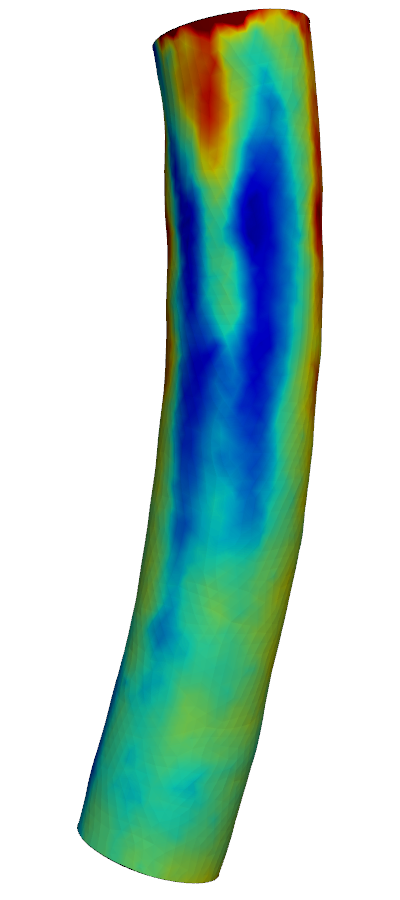} \\[-10pt]
         \rotatebox{90}{\parbox{3.0cm}{\centering Dataset 7}} & 
         \includegraphics[width=0.2\linewidth]{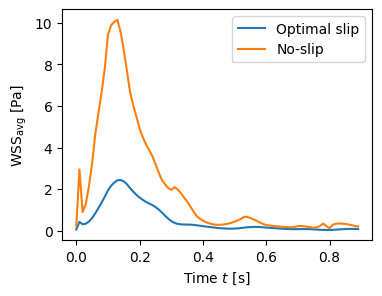} & 
         \includegraphics[height=0.1\textheight]{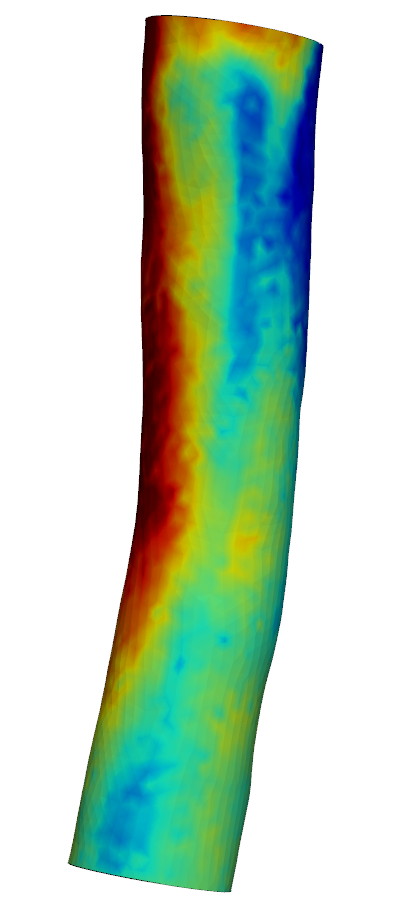} & 
         \includegraphics[height=0.1\textheight]{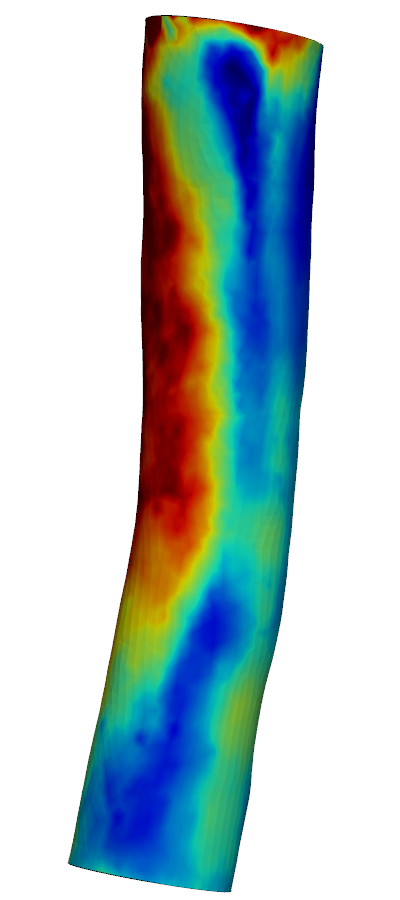}
    \end{tabular}  
    \caption{Comparison of wall shear stress (WSS) for optimal slip (blue line) and no-slip (orange line) conditions across all datasets using manual segmentation. Each row corresponds to one dataset: average WSS over time (left), WSS snapshot at $t=0.2$\,s with optimal slip (middle), and with no-slip (right). The no-slip condition produces two to three times higher WSS than the optimal slip condition.}
    \label{fig:WSS}
\end{figure}

The analysis reveals markedly lower (1/3 to 1/2) WSS magnitudes in the realistically slipping flow than in the unphysical no-slip case, see snapshots of wall shear stress at $t=0.2$\,s, near peak systole, and the time evolution of average WSS in \Cref{fig:WSS}. This is due to smaller velocity gradient near the wall, caused by partial slip.

\section{Discussion}
The scientific study of blood flow and its regulation dates back several centuries, and a mathematical approach to it---motivating the development of fluid mechanics broadly---traces at least back to Bernoulli and Euler in the early-to-mid 18th century \cite{DARRIGOL2002}. Early experimental studies in fluid mechanics revealed that flow characteristics differ significantly between the free-stream region and the area near the boundary. This observation led to Stokes' formulation of the `no-slip' condition in 1845 \cite{Stokes1845}, though he recognized that it may not always hold, especially for faster flows. Although some experimental evidence suggests that blood exhibits slip \emph{in vitro}, whether this occurs \emph{in vivo} remains unknown. Nevertheless, the no-slip condition has been universally enforced in blood flow modeling studies, largely because it simplifies problem solving.

Our data assimilation method --- originally developed in \cite{steady} for steady flows with artificially generated data and extended here to the time-dependent flows using real subject data --- enables the determination of velocity fields with the slip parameter $\kappa_{\rm opt}$ (and the inlet velocity $\vin$) that best match the MRI data in the whole volume of interest. (It also yields an estimate of the pressure field, which cannot be directly measured.)

Based upon our review of the literature, this work appears to provide novel evidence of blood slippage against the arterial wall \emph{in vivo}. A variety of physiological factors, perhaps most notably wall shear stress, are altered by the presence of slip at the blood/artery interface.

The ascertained decrease in wall shear stress has important physiological implications. WSS regulates endothelial signaling, inflammation, and the initiation of aortic dissection and atherosclerotic plaque formation. Our findings therefore challenge a central assumption of cardiovascular fluid mechanics.

Lower WSS may profoundly affect endothelial mechanotransduction. This may alter effector responses, e.g., nitric oxide production and local vascular permeability. Lower WSS may also influence platelet activation and thrombus formation, processes strongly dependent on shear rate. In the context of mathematical modeling, the presence of slip changes predicted pressure drops, vorticity distributions, and energy dissipation, thereby modifying all downstream quantities derived from simulations that assume no-slip. Together, these effects emphasize that an accurate representation of boundary behavior is essential for realistic modeling of cardiovascular flow and pathology.

\section{Conclusion}
Boundary phenomena in arteries are of vital importance. The vast majority of acute and chronic cardiovascular pathologies occur at boundaries. These include aortic dissection, aneurysm rupture, and atherosclerotic plaque formation across a large range of arterial beds – coronary, carotid, and peripheral -- with a risk of plaque rupture or organ hypoperfusion. A sound understanding of the behavior of blood at the wall is therefore essential, since any mathematical model of blood flow necessarily relies on a specific choice of boundary conditions. Accurate mathematical models combined with patient-specific information may have predictive capacity \cite{Candreva2022,Schwarz2023}.
Based upon review of the extant literature, {\bf this work appears to be the first study of slip conditions in arterial blood flow \emph{in vivo}}, and more specifically, demonstrates consistent failure of the no-slip condition to hold true. This may have significant implications for the study of a range of cardiovascular diseases.

\begin{acknowledgments}
Jaroslav Hron, Alena Jarolímová, Josef Málek and Karel Tůma have been supported by the project No. 23-05207S financed by the Czech Science Foundation, Czech Republic (GAČR). Their work has also been supported by Charles University Research Centre, Czech Republic program No. UNCE/24/SCI/005. J.H., J.M., and K.T. are members of the Nečas Center for Mathematical Modeling. The authors are also thankful for sharing the MRI datasets previously acquired within the British Heart Foundation project “Integrated Mathematical Modelling and Imaging for Dialated Cardiomyopathy” (2013 – 2014) by Myrianthi Hadjicharalambous, Liya Asner, Eva Sammut, RC, David Nordsletten (King’s College London, UK).\\[2mm] 
\end{acknowledgments}

\section*{Author Contributions}
All authors contributed equally to this work.

\section*{Data availability statement}
The source code for this study is publicly accessible on GitHub:
\url{https://github.com/jarolimova/Determination-of-Navier-slip-using-data-assimilation-using-Firedrake}

\end{document}